\documentclass[oneline]{IET}

\usepackage{multirow}
\usepackage{multicol}
\usepackage{ctable}
\usepackage{makecell}
\usepackage{boldline}
\usepackage{cite}
\usepackage{graphicx}
\usepackage{subfigure}
\newcolumntype{C}[1]{>{\centering\let\newline\\\arraybackslash\hspace{0pt}}m{#1}}

\graphicspath{{fig/}}

\begin{document}

\title{A Joint  Optimization Technique for Multi-Edge Type LDPC  Codes}

\author{Sachini Jayasooriya}
\author{Mahyar Shirvanimoghaddam}
\author{Lawrence Ong}
\author{Sarah J. Johnson}

\affil{School of Electrical Engineering and Computer Science, The University of Newcastle, Newcastle, Australia}
\affil{E-mail:c3192163@uon.edu.au}

\abstract{This paper considers the optimization of   multi-edge type low-density parity-check (MET-LDPC) codes to maximize the decoding threshold. We propose an  algorithm  to jointly optimize  the node degree distribution  and the multi-edge structure of MET-LDPC codes for  given values of the maximum  number of edge-types and maximum node degrees. This joint optimization   is particularly important for MET-LDPC codes as it is not clear a priori which structures will be good. Using several examples, we demonstrate that the MET-LDPC codes designed by the proposed joint optimization algorithm  exhibit improved decoding thresholds compared to previously reported MET-LDPC codes.}

\maketitle

\section{Introduction} \label{Introduction}

The field of error correcting codes was revolutionized with the introduction of graph-based codes such as turbo~\cite{BerrouNearICC1993}, low-density parity-check (LDPC)~\cite{GallagerLow-densityIT1962} and repeat-accumulate~\cite{JinIrregularTurbo2000} codes, and  much progress has been made towards understanding the performance of these codes, as well as designing even better codes with improved decoding performances.   These graph-based codes can be efficiently represented by a  Tanner graph in which the variable nodes correspond to the elements of the codeword and the check nodes to the parity-check constraints~\cite{TannerRecursiveIT1981}.  The LDPC codes drew a lot of attention to themselves due to their impressive performance over  standard channels such as the binary input additive white Gaussian noise (BI-AWGN) channel~\cite{RichardsonDesignIT2001,chung2001design}. 
Multi-edge type low-density parity-check (MET-LDPC) codes,  introduced by Richardson and Urbanke~\cite{RichardsonMulti2002}, provide a very powerful and general framework for the standard LDPC codes.  Unlike standard LDPC codes which contain a single statistical equivalence types of Tanner graph edges, in the MET setting several edge-types can be defined and every node is characterized by the number of connections to edges of each edge-type. This feature of the MET-LDPC code gives rise to Tanner graph structures not possible in the standard LDPC framework and  allows for exploring   codes, optimized under specific constraints, with better decoding performances~\cite{RichardsonMulti2002}.

In this paper, efficient design of  MET-LDPC codes on the binary erasure channel (BEC) and the BI-AWGN channel is considered.  A key performance measure of a  code ensemble is its decoding threshold. A code ensemble is the set of all  codes with a particular property set, usually the degree distribution of their Tanner graph representation, where  the degree distribution~\cite{LubyImprovedIT2001} (in either node perspective or edge perspective) specifies the connectivity among the nodes in the Tanner graph.  The decoding threshold of a given degree distribution is the  ``worst'' channel parameter  for which the  decoding will be successful with probability approaching one (i.e., decoding error probability tends to zero)   as the code length tends to infinity. The relevant channel parameters for the BEC and the BI-AWGN channel are, respectively, the erasure probability and the variance of the Gaussian noise.  A numerical technique, called Density Evolution (DE), was formulated to find the decoding threshold of the belief propagation (BP) decoding algorithm for a given LDPC~\cite{LubyImprovedIT2001,RichardsonDesignIT2001,richardson2001capacity} or MET-LDPC~\cite{RichardsonModernBook2008} code ensemble. DE determines the performance of the BP decoding for a given code ensemble by iteratively tracking the probability density function of messages passed along the edges in the corresponding Tanner graph. Then it is possible to test whether  the decoder can successfully decode the transmitted message for the given channel condition.  By running DE for different channel parameters, we can determine the decoding threshold. This allows  us to design and optimize LDPC and MET-LDPC code ensembles using the decoding threshold  as the cost function.

The code optimization of  LDPC and MET-LDPC codes is a non-linear cost function maximization problem, where the decoding threshold is the cost function and the Tanner graph structure (i.e., connection of edge-types and allowed degree set) and  degree distribution are the variables to be optimized. Efficient design of LDPC and MET-LDPC codes using different optimization methods   have been widely investigated in the literature~\cite{RichardsonMulti2002,RichardsonDesignIT2001,azmi2011design,ShokrollahiDesignBook2005,Tavakoli2012Optimal,Jamali2015Design,Luby2001efficinet}.
In the majority of previous work concerning code optimization for the MET-LDPC codes~\cite{RichardsonMulti2002,azmi2011design}, the Tanner graph structure is determined via trail and error or exhaustive search,  while only the  degree distribution within a given structure is optimized. For the LDPC codes, it has been found that the structure  of a very good  code  consists of  two or three non-zero check node degrees (which may be chosen consecutively) and variable nodes with  degrees two and three plus a very high degree  and some two or three degrees in between~\cite{MoonErrorBOOK2005}. However, it is not yet clear what structure constitutes a ``very good'' MET-LDPC code. The most common  method used to optimize the  degree distribution of LDPC and MET-LDPC codes is  Differential Evolution (Dif.E)~\cite{storn1997differential1997,ShokrollahiDesignBook2005}.  Some other  methods used to finding good degree distributions are hill climbing~\cite{RichardsonMulti2002}, genetic algorithms~\cite{RichardsonDesignIT2001} and linear programming~\cite{Luby2001efficinet}. However, these methods suffer from some disadvantages, such as not guaranteeing a feasible solution and being sensitive to their starting point. 

In this work we  first develop a new code optimization technique to optimize MET-LDPC codes more efficiently, particularly the degree distribution. This technique can be thought of as minimizing the randomness in Dif.E~\cite{storn1997differential1997,ShokrollahiDesignBook2005} or limiting the search space in ordinary exhaustive search and hill-climbing~\cite{RichardsonMulti2002}. We then propose a joint optimization method to optimize both the  overall  Tanner graph structure  and the degree distribution of  MET-LDPC codes given the number of edge-types and maximum node degrees. This allows us to systematically find good  MET-LDPC  code structures.  Unlike LDPC codes, for which we roughly know what structures are sufficiently good, it is not  clear a priori what structures will be good for the MET-LDPC codes. Thus the joint optimization  is particularly important to design good  MET-LDPC codes. We then use our proposed joint optimization method to design good  MET-LDPC codes  with various rates over the BEC and the BI-AWGN channel.

This paper is organized as follows. Section~\ref{Background} briefly reviews some basic concepts of   MET-LDPC codes.   In Section~\ref{optimization MET-LDPC}, we present our proposed methodology and in Section~\ref{discussion}  we discuss the code optimization results obtained for several  MET-LDPC codes. Finally, Section~\ref{Conclusion} concludes the paper.

\section{Background of MET-LDPC codes } \label{Background}

%\subsection{MET-LDPC code ensemble} \label{MET-LDPC codes}
An MET-LDPC code ensemble can be specified  by a pair of node-perspective degree distribution polynomials related to the variable nodes and check nodes respectively~{\protect\cite[page 383]{RichardsonModernBook2008}}:
\vspace*{-1em}
\begin{align}
L(\boldsymbol{r},\boldsymbol{x}) &= \sum_{(\boldsymbol{b},\boldsymbol{d})} L_{\boldsymbol{b},\boldsymbol{d}}~ \boldsymbol{r}^{\boldsymbol{b}}~ \boldsymbol{x}^{\boldsymbol{d}}			\label{eq : MET_LDPC_lamda}\\
R(\boldsymbol{x}) &= \sum_{\boldsymbol{d}} R_{\boldsymbol{d}} ~\boldsymbol{x}^{\boldsymbol{d}}			
\label{eq : MET_LDPC_roh}
\end{align}

In the graph representation of MET-LDPC codes, the nodes are categorized    into different node classes.  A node-class is the partition of the nodes  such that all the nodes in the class have the same property (e.g., degree). Let a variable node class be identified by $(\boldsymbol{b},\boldsymbol{d})$, and a check node class by $(\boldsymbol{d})$.   $L_{\boldsymbol{b},\boldsymbol{d}}$ and $R_{\boldsymbol{d}}$  correspond to the fraction of variable nodes with type ($\boldsymbol{b}, \boldsymbol{d}$)  and the fraction of check nodes with type ($\boldsymbol{d}$),  respectively, of a Tanner graph in the ensemble. Each term in the summation of (\ref{eq : MET_LDPC_lamda}) describes a variable node class (of type ($\boldsymbol{b}, \boldsymbol{d}$)), and each term in the summation of (\ref{eq : MET_LDPC_roh}) a check node class (of type ($\boldsymbol{d}$)). %Using this characterization, $[(\boldsymbol{b}, \boldsymbol{d})]$ and $[(\boldsymbol{d})]$ define the Tanner graph structure, and $[L_{\boldsymbol{b},\boldsymbol{d}}]$ and $[R_{\boldsymbol{d}}]$ the degree distribution. 

Let $m_e$ denote the number of edge-types corresponding to the Tanner graph.   The vector $\boldsymbol{d}=[d_1 ~\dots ~d_{m_e}]$ indicates the degrees of each node in the class and we use $\boldsymbol{x}^{\boldsymbol{d}}$ to denote $\prod_{i=1}^{m_e} x_i^{d_i}$. Each node of type $(-,\boldsymbol{d})$ or $(\boldsymbol{d})$ has $d_i$ number of edges of edge-type $i$ connected to it. For example, a node associated with $x_1^3x_2^5$ is connected to  three edges of  edge-type $1$, and five edges of  edge-type $2$. Unlike standard LDPC codes, the MET framework allows variable nodes to receive information from different channels.   Thus there is an additional vector,  $ \boldsymbol{r}^{\boldsymbol{b}} = \prod_{i=0}^{m_r} r_i^{b_i}$, associated with each variable node, where $\boldsymbol{b}$ indicates the channel to which the variable node is connected and  $m_r$ denotes the number of different channels over which a codeword bit may be transmitted. Typically for binary input channels, $\boldsymbol{b} = [b_0 ~\dots ~b_{m_r}]$,  has only two entries as a codeword bit is either punctured (the codeword bits not transmitted: $\boldsymbol{b} = [1~0]$ and $\boldsymbol{r}^{\boldsymbol{b}}=r_0$) or transmitted through a single channel ($\boldsymbol{b} = [0~1]$ and $\boldsymbol{r}^{\boldsymbol{b}}=r_1$). For example, $r_0x_1^3x_2^5$ is interpreted as a punctured variable node and $r_1x_1^3x_2^5$ as a variable node connected to channel $1$.  

The code rate of a MET-LDPC code ensemble is given by~{\protect\cite[page 383]{RichardsonModernBook2008}},
\vspace*{-1em}
\begin{align}
	\mathcal{R} &= L(\boldsymbol{1},\boldsymbol{1}) - R(\boldsymbol{1}).
	\label{code rate}
\end{align}
where $\boldsymbol{1}$ denotes a vector of all $1$'s with the length determined by the context.

Throughout this paper,  we use $\boldsymbol{\Lambda}=[(\boldsymbol{b}_{v_1},\boldsymbol{d}_{v_1}) ~(\boldsymbol{b}_{v_2},\boldsymbol{d}_{v_2}) \dots (\boldsymbol{b}_{v_m},\boldsymbol{d}_{v_m})]$ and 
$\boldsymbol{\Gamma}=[\boldsymbol{d}_{c_1} ~\boldsymbol{d}_{c_2} \dots \boldsymbol{d}_{c_m}]$ to denote the Tanner graph structure for  the variable nodes and the check nodes, respectively, where $v_m$ is the number of variable node classes and $c_m$ is the number of check node classes. $(\boldsymbol{b}_i,\boldsymbol{d}_i), i=v_1,v_2,\dots,v_m$, is the variable node class identified by type $(\boldsymbol{b}_i,\boldsymbol{d}_i)$ and $\boldsymbol{d}_j, j=c_1,c_2,\dots,c_m$, is the check node class identified by type $(\boldsymbol{d}_j)$.  Moreover, we use $\boldsymbol{L}= [L_{\boldsymbol{b}_{v_1},\boldsymbol{d}_{v_1}} ~L_{\boldsymbol{b}_{v_2},\boldsymbol{d}_{v_2}} \dots L_{\boldsymbol{b}_{v_m},\boldsymbol{d}_{v_m}}]$ and $\boldsymbol{R}=[R_{\boldsymbol{d}_{c_1}} ~R_{\boldsymbol{d}_{c_2}} \dots R_{\boldsymbol{d}_{c_m}}]$ to denote  the degree distribution of the variable nodes and the check nodes, respectively. The order in $\boldsymbol{\Lambda}, \boldsymbol{\Gamma}, \boldsymbol{L}, ~\text{and} ~\boldsymbol{R}$ is arbitrary but fixed.

%figure for example MET-LDPC tanner graph
\begin{figure}[t!]
	\centering
	\includegraphics[width=0.5\linewidth]{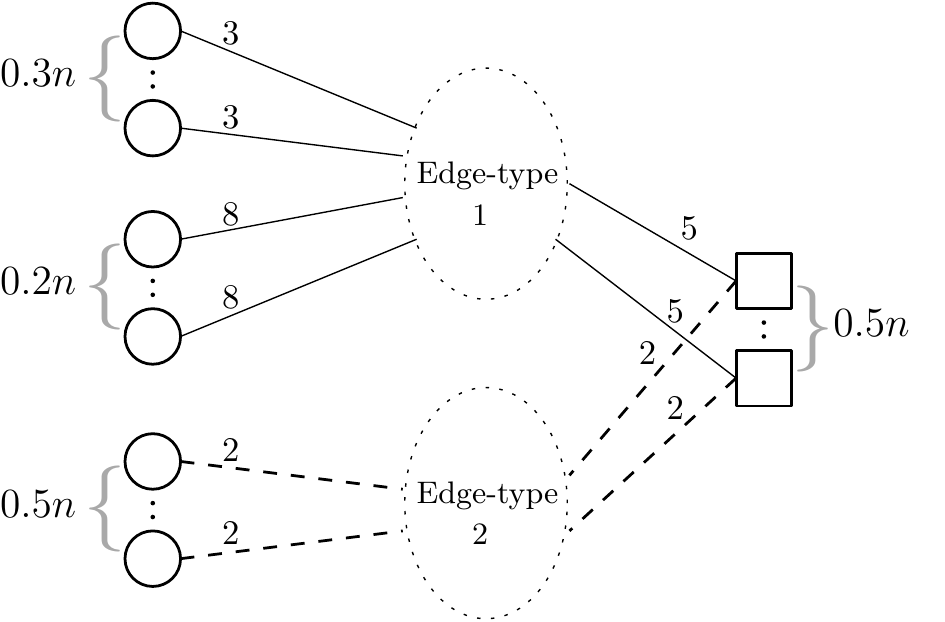}
	\caption{Graphical representation of an example two edge-type MET-LDPC code (the rate-$1/2$ DVB code~\cite{DVB}), where \textquoteleft $\circ$\textquoteright~   represents   variable nodes and \textquoteleft $\Box$\textquoteright~ represents check nodes. The number of nodes for different edge-types are shown as fractions of the code length $n$, where $n$ is the number of transmitted code bits.}
	\label{fig:MET_Tanner}
\end{figure}

A rate-$1/2$  MET-LDPC code ensemble is shown in Fig.~\ref{fig:MET_Tanner}, where the node-perspective degree distribution is given by $L(\boldsymbol{r},\boldsymbol{x}) = 0.3r_1x_1^3+0.2r_1x_1^8+0.5r_1x_2^2$ and $R(\boldsymbol{x}) = 0.5x_1^5x_2^2$. This can  alternatively  be represented as: $\boldsymbol{\Lambda} = [([0~1], [3 ~0]) \hspace*{0.3cm}([0~1], [8 ~0]) \hspace*{0.3cm}([0~1], [0 ~2])]$, $\boldsymbol{\Gamma} = [5 ~2]$, $\boldsymbol{L} = [0.3  \hspace*{0.3cm}0.2  \hspace*{0.3cm}0.5]$ and $\boldsymbol{R} = [0.5]$.

%\begin{remark}
%	We can remark that the standard LDPC codes can be considered as a special case of MET-LDPC codes which has a single edge-type (i.e., $m_e=1$) and all the codeword bits are transmitted over the same channel (i.e., $m_r=1$). We can consider the rate-$1/2$  MET-LDPC code  shown in Fig.~\ref{fig:MET_Tanner} as a rate-$1/2$ standard LDPC code (i.e., single edge-type MET-LDPC code) with  $\Lambda$ = [3, 8, 2], $\Gamma$ =[7]. The edge-perspective degree distribution is given by $\lambda(x) = 0.3104x^2 + 0.5517x^7 + 0.1379x$  and $\rho(x) = x^6$.	 
%\end{remark}

\section{Design of MET-LDPC codes} \label{optimization MET-LDPC}

\subsection{Problem statement for MET-LDPC code optimization}	\label{problem statemet}

The code optimization of MET-LDPC codes is a non-linear cost function maximization problem where the decoding threshold (${P}^{*}$) is the cost function and the Tanner graph structure ($\boldsymbol{\Lambda}, \boldsymbol{\Gamma}$) and  degree distribution ($\boldsymbol{L}, \boldsymbol{R}$) give the variables to be optimized. Our goal is to optimally choose the elements of ($\boldsymbol{\Lambda}, \boldsymbol{\Gamma}, \boldsymbol{L}, \boldsymbol{R}$)  so that the corresponding code ensemble yields the largest possible threshold, which 
can be  determined  via DE. On the BEC the optimization problem is as follows.

For a fixed  code rate,  $\mathcal{R}\in [0, 1]$\\
$\max {P}^{*}(\boldsymbol{\Lambda}, \boldsymbol{\Gamma}, \boldsymbol{L}, \boldsymbol{R})$,  \\
Subject to 
\vspace*{-2em} 
\begin{align} 
%0 \leq L_{\boldsymbol{b},\boldsymbol{d}}, R_{\boldsymbol{d}} &\leq 1\\
\sum_{(\boldsymbol{b},\boldsymbol{d}):\boldsymbol{b}=[0~1]}L_{\boldsymbol{b},\boldsymbol{d}} &= 1  \label{c1}\\
L(\boldsymbol{1},\boldsymbol{1}) - R(\boldsymbol{1}) &= \mathcal{R} \label{c2}\\
L_{x_i}(\boldsymbol{1},\boldsymbol{1}) &= R_{x_i}(\boldsymbol{1}).  \label{c3}					
\end{align}

Here  ${P}^{*}(\boldsymbol{\Lambda}, \boldsymbol{\Gamma}, \boldsymbol{L}, \boldsymbol{R})$ can be computed using a BEC density evolution algorithm~\cite{JohnsonIterativeBook2009}. On other channels an appropriate DE function, or a suitable approximation, can be used to calculate ${P}^{*}(\boldsymbol{\Lambda}, \boldsymbol{\Gamma}, \boldsymbol{L}, \boldsymbol{R})$.

When optimizing the MET-LDPC code ensemble, there are several constraints that need to be satisfied.  (\ref{c1}) and (\ref{c2})  are imposed to satisfy the constraints on  the total number of transmitted code bits as fractions of the code length and the code rate, respectively.  One other constraint that needs to be satisfied is  the total  number of edges of each edge-type in variable node side and check node side. This is the socket count equality constraint that is given in (\ref{c3}), where
\vspace*{-1em}
\begin{align}
	L_{x_i}(\boldsymbol{1},\boldsymbol{1}) &= \frac{d}{d{x_i}} L(\boldsymbol{r},\boldsymbol{x})\biggr|_{\boldsymbol{r}=\boldsymbol{1},\boldsymbol{x}=\boldsymbol{1}} \\
	R_{x_i}(\boldsymbol{1}) &= \frac{d}{d{x_i}} R(\boldsymbol{x})\biggr|_{\boldsymbol{x}=\boldsymbol{1}}. 
\end{align}

Traditionally, the optimization problem is considered for fixed $\boldsymbol{\Lambda}$ and $\boldsymbol{\Gamma}$ chosen via trail and error or intuition. That is, the allowed degrees (for which $\boldsymbol{L}$ and $\boldsymbol{R}$ are non-zero) are fixed in advance. Here we include the allowed degrees as variables in the optimization which is particularly useful for the generalization of MET-LDPC codes where the choice of $\boldsymbol{\Lambda}$ and $\boldsymbol{\Gamma}$ is not straightforward. We called this joint optimization procedure a ``combined optimization''.  In the combine optimization,  we add some other constraints to reduce the search space  of ($\boldsymbol{\Lambda}, \boldsymbol{\Gamma}$) as follows:
\vspace*{-2em} 
\begin{align}
m_e &\leq m_{e_{\max}}\\ 
v_m &\leq v_{\max}\\
c_m &\leq c_{\max}\\
\max_{i \in \{v_1,\dots,v_m\}} |\boldsymbol{d}_i| &\leq d_{v_{\max}}
\end{align}
where $m_{e_{\max}}$ is the maximum number of edge-type, $v_{\max}$ (resp., $c_{\max}$) is the maximum number of variable node  classes (resp., check node classes) allowed.   $| \boldsymbol{d}_i| =d_1+d_2+\dots+d_{m_e}$ denotes the degree of the variable node class type $(\boldsymbol{b}_i,\boldsymbol{d}_i)$ and $d_{v_{\max}}$ is used to denote the maximum variable node  degree allowed. 

\subsection{Computation of $\boldsymbol{\Gamma}$ and $\boldsymbol{R}$ } \label{check node}

Before describing the optimization procedure,  we need to first identify the dependencies among the elements of ($\boldsymbol{\Lambda}, \boldsymbol{\Gamma}, \boldsymbol{L}, \boldsymbol{R}$). It has been shown~\cite{RichardsonMulti2002} that,  it is essential to satisfy  the code rate constraint and the socket count equality constraint in (\ref{c2}) and (\ref{c3}) in order to have a  valid MET-LDPC code ensemble. These two constraints allow us to include ($\boldsymbol{\Gamma}, \boldsymbol{R}$) as the dependent variables in the code optimization. Our  design methodology  involves  choosing  concentrated check node degrees based on the $\mathcal{R}$,  $\boldsymbol{\Lambda}$, and  $\boldsymbol{L}$. The procedure for generating concentrated check node degrees for rate-$\mathcal{R}$ two edge-type MET-LDPC code with a given variable node degree distribution, $L(\boldsymbol{r},\boldsymbol{x})$, (i.e., for a fixed $\boldsymbol{\Lambda}, \boldsymbol{L}$) is as follows.

\begin{enumerate}
\item{The total number of type-$1$ and type-$2$ edges are given by $L_{x_1}(\boldsymbol{1},\boldsymbol{1})$ and $L_{x_2}(\boldsymbol{1},\boldsymbol{1})$, respectively.  The average  check node degrees for the edge-type $1$ and $2$  can then be computed using (\ref{c2}) and (\ref{c3}) as follows. 
\vspace*{-1em}
\begin{align*}
{d}_{\text{avg}} &= \frac{R_{x_1}(\boldsymbol{1})}{R(\boldsymbol{1})} = \frac{L_{x_1}(\boldsymbol{1},\boldsymbol{1})}{L(\boldsymbol{1},\boldsymbol{1})-\mathcal{R}} \\
{\bar{d}}_{\text{avg}} &= \frac{R_{x_2}(\boldsymbol{1})}{R(\boldsymbol{1})} = \frac{L_{x_2}(\boldsymbol{1},\boldsymbol{1})}{L(\boldsymbol{1},\boldsymbol{1})-\mathcal{R}} 
\end{align*}}
\item{The coefficients of the concentrated  check node degrees  can be computed by applying socket count equality constraint (\ref{c3}) to edge-type $1$ as follows. $\lceil\cdot \rceil$ and $\lfloor\cdot \rfloor$ denote the ceiling and floor functions, respectively. 
\vspace*{-1em}		 
\begin{align*}
R_{\boldsymbol{d}_{c_1}} &= (L(\boldsymbol{1},\boldsymbol{1})-\mathcal{R})\cdot\lceil {d}_{\text{avg}}\rceil - L_{x_1}(\boldsymbol{1},\boldsymbol{1}) \\
R_{\boldsymbol{d}_{c_2}} &= (L(\boldsymbol{1},\boldsymbol{1})-\mathcal{R})\cdot(1-\lceil {d}_{\text{avg}}\rceil) + L_{x_1}(\boldsymbol{1},\boldsymbol{1})  		 
\end{align*}
Similarly, we can apply socket count equality constrain  (\ref{c3}) to edge-type $2$ to compute the coefficients of the concentrated  check node degrees, which gives
\vspace*{-1em}
\begin{align*}
\bar{R_{\boldsymbol{d}_{c_1}}} &= (L(\boldsymbol{1},\boldsymbol{1})-\mathcal{R})\cdot\lceil {\bar{d}}_{\text{avg}}\rceil - L_{x_2}(\boldsymbol{1},\boldsymbol{1}) \\
\bar{R_{\boldsymbol{d}_{c_2}}} &= (L(\boldsymbol{1},\boldsymbol{1})-\mathcal{R})\cdot(1-\lceil {\bar{d}}_{\text{avg}}\rceil) + L_{x_2}(\boldsymbol{1},\boldsymbol{1})  		 
\end{align*}}
\end{enumerate}

\begin{figure}[!h]
 	\centering
 	\includegraphics[width=0.55\linewidth]{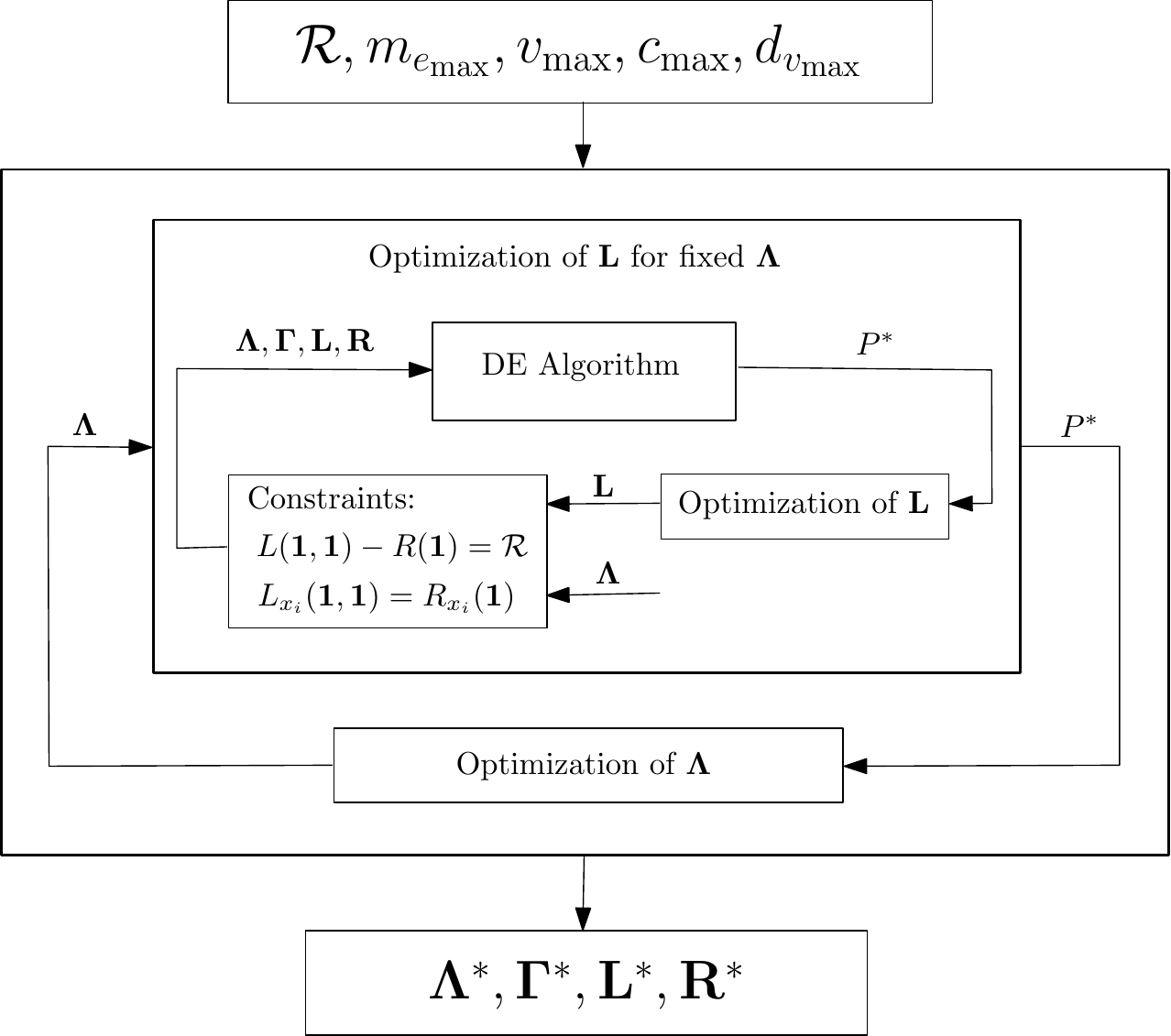}
 	\caption{ The block diagram for the combined optimization method}
 	\label{fig:block_dig}
 \end{figure}

\begin{list}{}{}
	\item {Case 1: Two check node classes\\
		If $R_{\boldsymbol{d}_{c_1}}=\bar{R_{\boldsymbol{d}_{c_1}}}$ and $R_{\boldsymbol{d}_{c_2}}=\bar{R_{\boldsymbol{d}_{c_2}}}$, we can compute the check node degree distribution using two check node classes. Then $\boldsymbol{\Gamma}$ and $\boldsymbol{R}$ are given by
		\vspace*{-1em}
		\begin{align*}
		\boldsymbol{\Gamma} &= [\boldsymbol{d}_{c_1} \hspace*{0.3cm} \boldsymbol{d}_{c_2}], \\	
		\text{where} \hspace*{0.3cm}\boldsymbol{d}_{c_1}&=[\lfloor {d}_{\text{avg}}\rfloor  \hspace*{0.3cm}\lfloor {\bar{d}}_{\text{avg}}\rfloor] \hspace*{0.3cm} \text{and} \hspace*{0.3cm} \boldsymbol{d}_{c_2}=[\lceil {d}_{\text{avg}}\rceil  \hspace*{0.3cm}\lceil {\bar{d}}_{\text{avg}}\rceil] \\
		\boldsymbol{R} &= [R_{\boldsymbol{d}_{c_1}} \hspace*{0.3cm} R_{\boldsymbol{d}_{c_2}}]							
		\end{align*}}
	\item {Case 2: Three check node classes\\
		If $R_{\boldsymbol{d}_{c_1}}\neq\bar{R_{\boldsymbol{d}_{c_1}}}$ and $R_{\boldsymbol{d}_{c_2}}\neq\bar{R_{\boldsymbol{d}_{c_2}}}$, we need three check node classes to  compute the check node degree distribution. Then $\boldsymbol{\Gamma}$ and $\boldsymbol{R}$ are computed as follows.\\
		If $\bar{R_{\boldsymbol{d}_{c_1}}} < R_{\boldsymbol{d}_{c_1}}$:
		\vspace*{-1em}
		\begin{align*}
		\boldsymbol{\Gamma} &= [\boldsymbol{d}_{c_1} \hspace*{0.3cm}\boldsymbol{d}_{c_2} \hspace*{0.3cm}\boldsymbol{d}_{c_3}],  \\
		\text{where} \hspace*{0.3cm} \boldsymbol{d}_{c_1}&=[\lfloor {d}_{\text{avg}}\rfloor  \hspace*{0.3cm}\lfloor {\bar{d}}_{\text{avg}}\rfloor], \hspace*{0.3cm}
		\boldsymbol{d}_{c_2}=[\lfloor {d}_{\text{avg}}\rfloor  \hspace*{0.3cm}\lceil {\bar{d}}_{\text{avg}}\rceil] \hspace*{0.3cm} \text{and} \hspace*{0.3cm}
		\boldsymbol{d}_{c_3}=[\lceil {d}_{\text{avg}}\rceil  \hspace*{0.3cm}\lceil {\bar{d}}_{\text{avg}}\rceil]\\	
		\boldsymbol{R} &= [\bar{R_{\boldsymbol{d}_{c_1}}} \hspace*{0.3cm}(R_{\boldsymbol{d}_{c_1}}-\bar{R_{\boldsymbol{d}_{c_1}}}) \hspace*{0.3cm}R_{\boldsymbol{d}_{c_2}}]							
		\end{align*}
		If $\bar{R_{\boldsymbol{d}_{c_1}}} > R_{\boldsymbol{d}_{c_1}}$:
		\vspace*{-1em}
		\begin{align*}
		\boldsymbol{\Gamma} &= [\boldsymbol{d}_{c_1} \hspace*{0.3cm}\boldsymbol{d}_{c_2} \hspace*{0.3cm}\boldsymbol{d}_{c_3}],  \\
		\text{where} \hspace*{0.3cm} \boldsymbol{d}_{c_1}&=[\lfloor {d}_{\text{avg}}\rfloor  \hspace*{0.3cm}\lfloor {\bar{d}}_{\text{avg}}\rfloor], \hspace*{0.3cm} 
		\boldsymbol{d}_{c_2}=[\lceil {d}_{\text{avg}}\rceil  \hspace*{0.3cm}\lfloor {\bar{d}}_{\text{avg}}\rfloor] \hspace*{0.3cm} \text{and} \hspace*{0.3cm}
		\boldsymbol{d}_{c_3}=[\lceil {d}_{\text{avg}}\rceil  \hspace*{0.3cm}\lceil {\bar{d}}_{\text{avg}}\rceil]\\	
		\boldsymbol{R} &= [R_{\boldsymbol{d}_{c_1}} \hspace*{0.3cm}(\bar{R_{\boldsymbol{d}_{c_1}}}-R_{\boldsymbol{d}_{c_1}}) \hspace*{0.3cm}\bar{R_{\boldsymbol{d}_{c_2}}}]							
		\end{align*}}
\end{list}
\vspace*{-1em}
We see that ($\boldsymbol{\Gamma}, \boldsymbol{R}$) is fixed given ($\boldsymbol{\Lambda}, \boldsymbol{L}$). Hence the optimization of ($\boldsymbol{\Lambda}, \boldsymbol{\Gamma}, \boldsymbol{L}, \boldsymbol{R}$) amounts to searching for the best ($\boldsymbol{\Lambda}, \boldsymbol{L}$).  Fig.~\ref{fig:block_dig} shows the block diagram for the combined optimization of MET-LDPC codes.
 
\subsection{The cost function of  MET-LDPC code optimization}

\begin{figure}[!b]
	\centerline{
		\subfigure[]{\includegraphics[width=3in ]{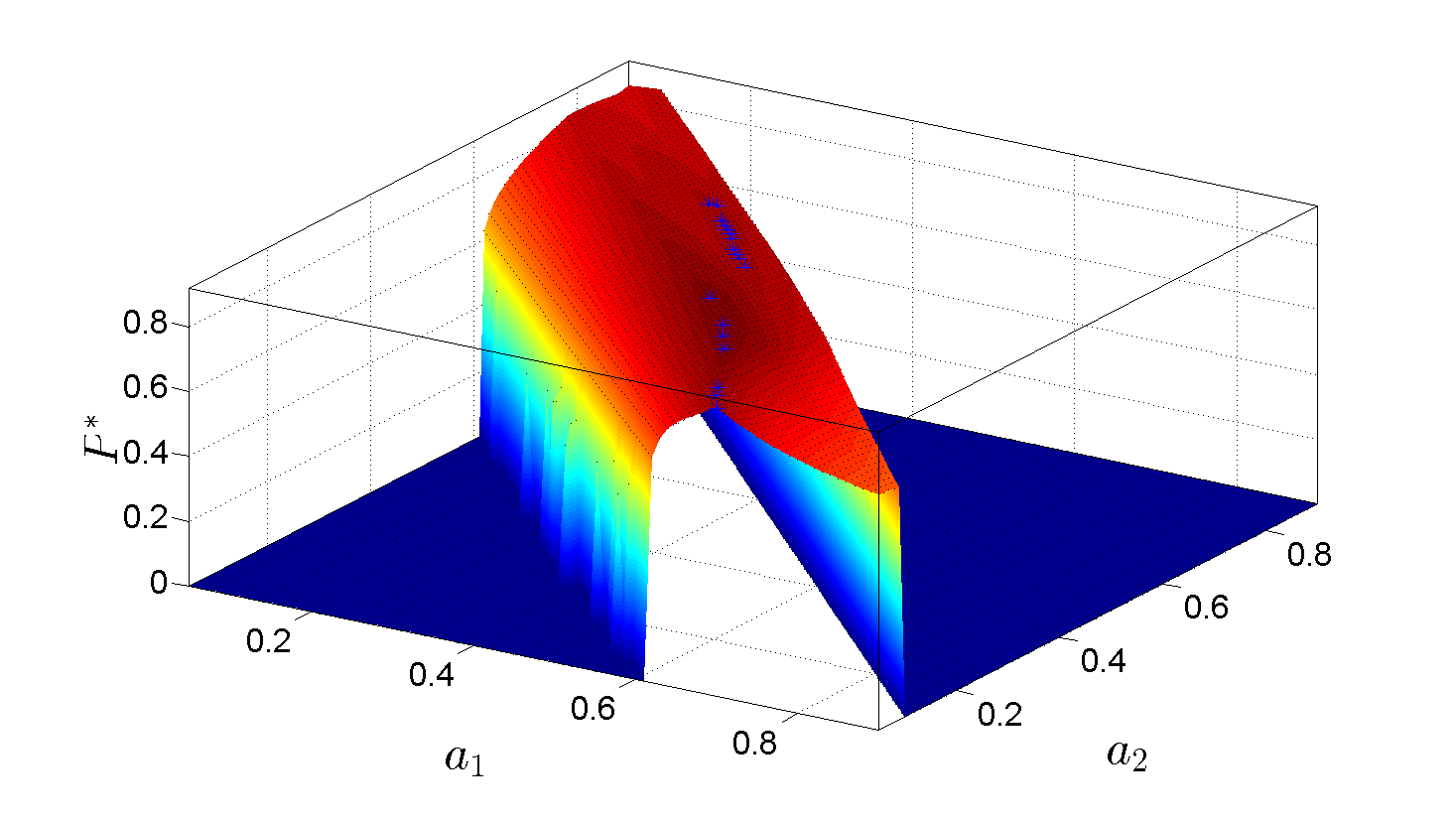} \label{rate_1_2_AWGN_inner}}  
		\hspace{1em}
		\subfigure[]{\includegraphics[width=3in]{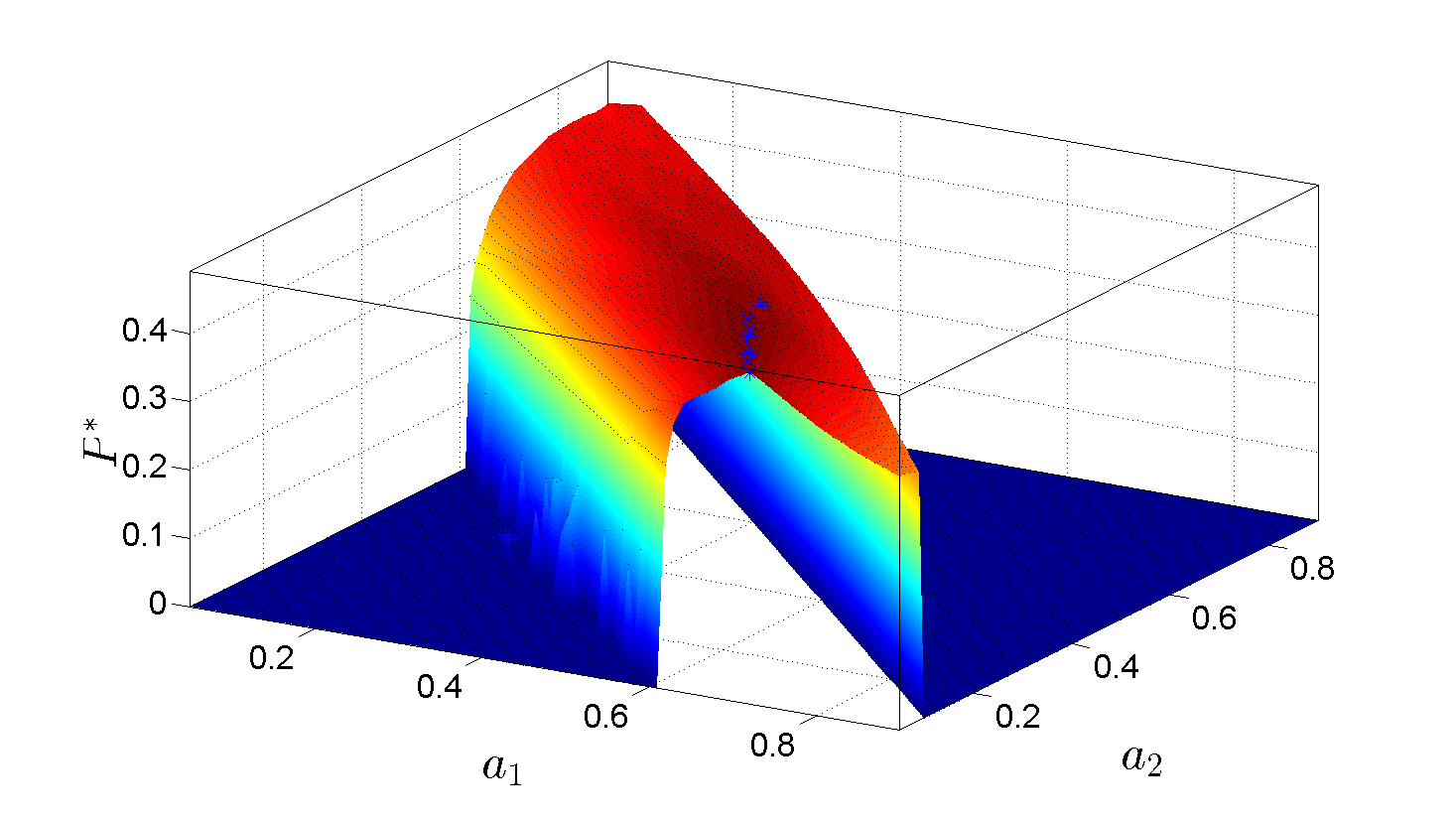} \label{rate_1_2_BEC_inner}} }
	\caption{The cost function of rate-$1/2$ MET-LDPC code for optimizing the  degree distribution, $\boldsymbol{L}$, for fixed, $\boldsymbol{\Lambda}$. $\boldsymbol{L}=[a_1 ~a_2 ~a_3 ~a_4]$  where $a_4 = 1-a_1-a_2$  and $a_3= a_4$ to satisfy constraint (\ref{c1}). Blue colour \textquoteleft \textasteriskcentered \textquoteright  ~shows the locations of local maxima.}
	\fontsize{10}{1}
	\subcaption{(a)}{On the BI-AWGN channel}
	\subcaption{(b)}{On the BEC}
	\label{inner_cost_rate_1_2}
\end{figure}

\begin{figure}[!b]
	\centerline{
		\subfigure[]{\includegraphics[width=3in ]{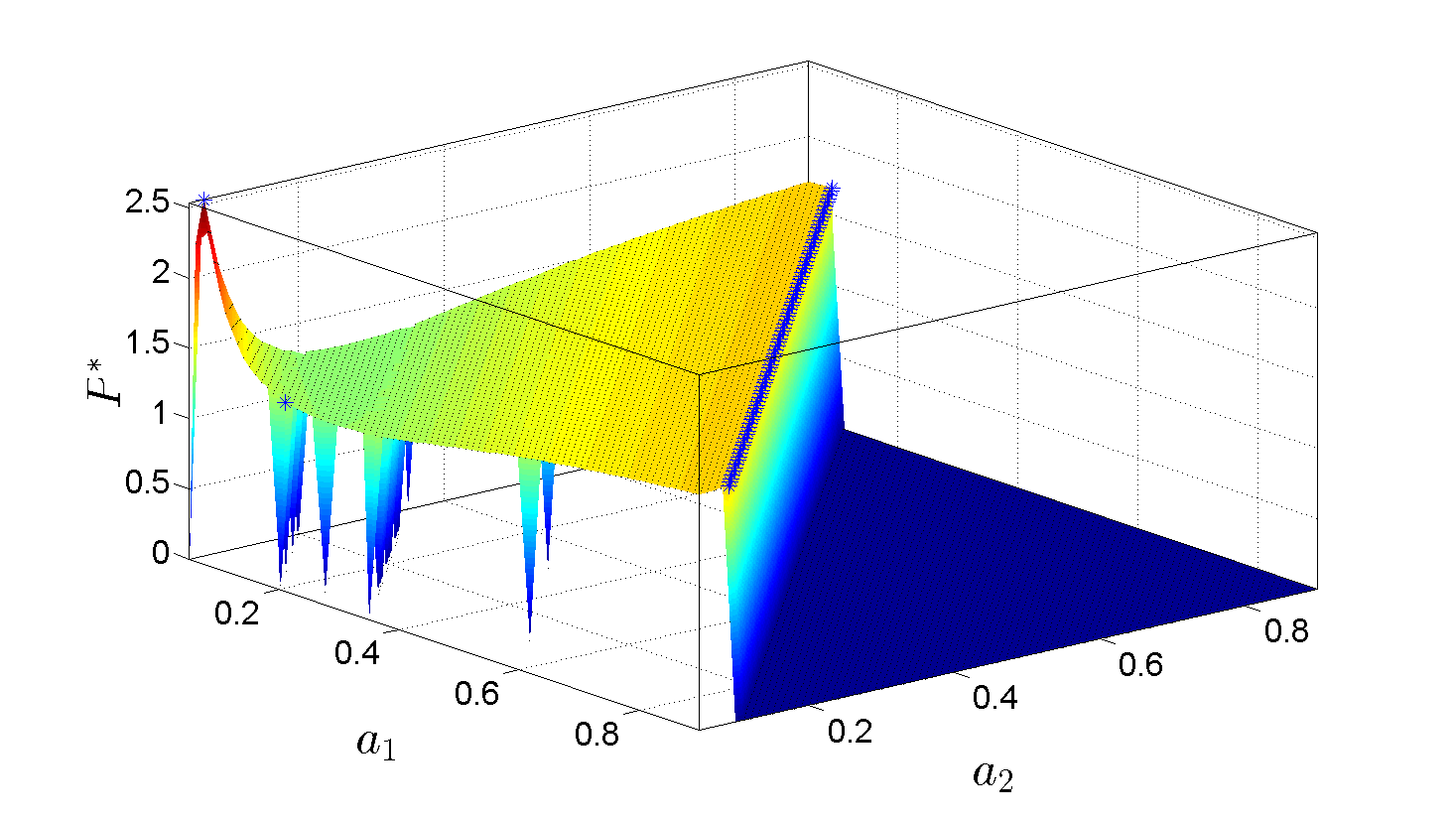} \label{rate_1_10_AWGN_inner}} 
		\hspace{1em}
		\subfigure[]{\includegraphics[width=3in]{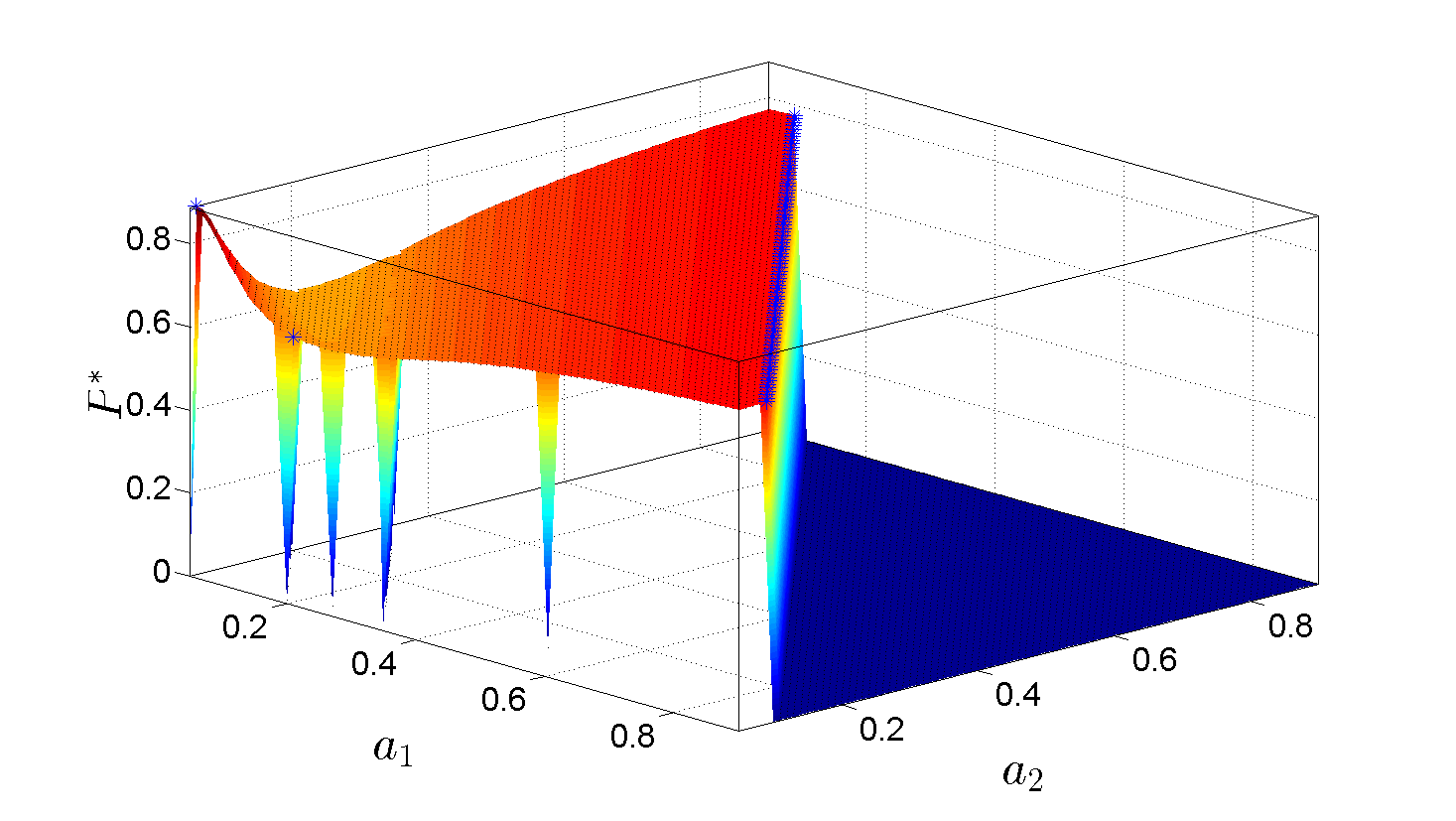}\label{rate_1_10_BEC_inner}}}
	\caption{The cost function of rate-$1/10$ MET-LDPC code for optimizing the  degree distribution, $\boldsymbol{L}$, for fixed, $\boldsymbol{\Lambda}$. $\boldsymbol{L}=[a_1 ~a_2 ~a_3]$  where $a_3 = 1-a_1-a_2$ to satisfy constraint (\ref{c1}). Blue colour \textquoteleft \textasteriskcentered \textquoteright  ~shows the locations of local maxima.}
	\fontsize{10}{1}
	\subcaption{(a)}{On the BI-AWGN channel}
	\subcaption{(b)}{On the BEC}
	\label{inner_cost_rate_1_10}
\end{figure}

As  explained in above sections, the combined optimization method consists of two stages. i.e., the optimization of the degree distribution, $\boldsymbol{L}$,  and the optimization of the Tanner graph structure, $\boldsymbol{\Lambda}$. Due to the different nature of $\boldsymbol{L}$ and $\boldsymbol{\Lambda}$, it could be useful to employ different optimization techniques for these two stages. In this section, we examine the properties of the cost function of  MET-LDPC code optimization (i,e., the decoding threshold (${P}^{*}$))
for optimizing $\boldsymbol{L}$ and   $\boldsymbol{\Lambda}$. We consider  rate-$1/2$ and rate-$1/10$ MET-LDPC code ensembles on the   BEC and  the BI-AWGN channel. The cost function, ${P}^{*}$, is evaluated using the BEC density evolution algorithm~\cite{JohnsonIterativeBook2009} and the DE approximation based on bit error rate (BER)~\cite{LehmannAnalysisIT2003} for the BEC and the  BI-AWGN channel, respectively. 

We first consider the cost function, ${P}^{*}$, for optimizing the degree distribution, $\boldsymbol{L}$, for a fixed Tanner graph structure, $\boldsymbol{\Lambda}$. For this case we use the MET-LDPC Tanner graph structures, proposed in the literature, for rate-$1/2$~{\protect\cite[Table VI]{RichardsonMulti2002}}  and rate-$1/10$~{\protect\cite[Table X]{RichardsonMulti2002}} MET-LDPC codes.  The  cost function for MET-LDPC degree distribution, when $\boldsymbol{\Lambda}$ is fixed, has local maxima. However they are close each other and concentrated into a small area.  An example of this  cost functions  on the BI-AWGN channel and the BEC are shown in Figs.~\ref{inner_cost_rate_1_2} and~\ref{inner_cost_rate_1_10}, as the coefficients of variable node class $1$ and $2$ are varied. 

\begin{figure}[!b]
	\centering
	\includegraphics[width=0.45\linewidth]{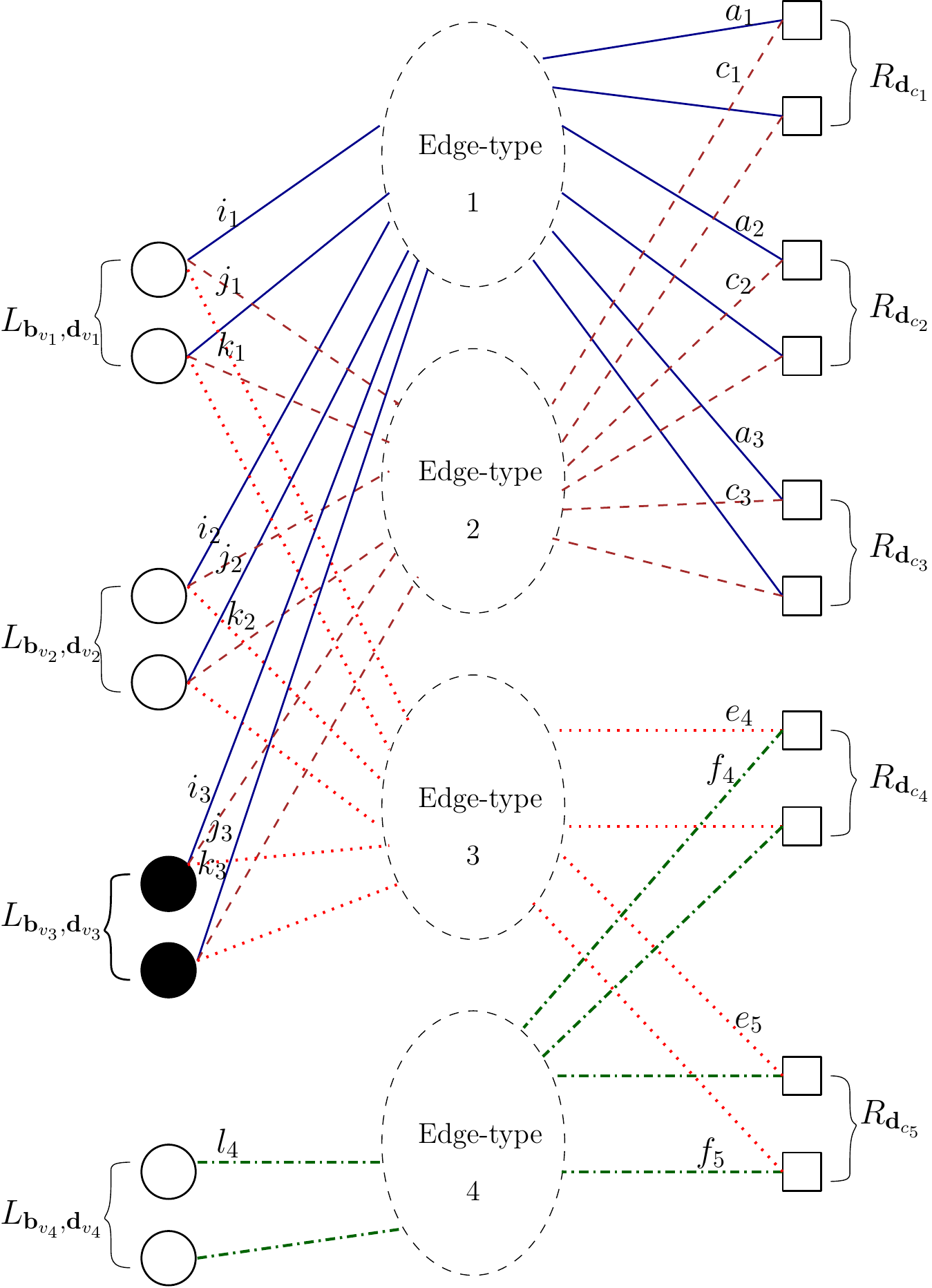}
	\caption{Tanner graph representation of four edge-type MET-LDPC code where subscript  letters denote the node class and ~\textquoteleft $\bullet$\textquoteright ~represents punctured variable nodes. $\boldsymbol{\Lambda} = [([0~1], [i_1 ~j_1 ~k_1 ~0]) \hspace*{0.3cm} ([0~1], [i_2 ~j_2 ~k_2 ~0]) \hspace*{0.3cm} ([1~0], [i_3 ~j_3 ~k_3 ~0]) \hspace*{0.3cm} ([0~1], [0 ~0 ~0 ~l_4])]$, $\boldsymbol{\Gamma} = [[a_1 ~c_1 ~0 ~0] \hspace*{0.3cm} [a_2 ~c_2 ~0 ~0] \hspace*{0.3cm} [a_3 ~c_3 ~0 ~0] \hspace*{0.3cm} [~0 ~0 ~e_4 ~f_4] \hspace*{0.3cm} [~0 ~0 ~e_5 ~f_5] ]$, $\boldsymbol{L} = [L_{\boldsymbol{b}_{v_1},\boldsymbol{d}_{v_1}}  \hspace*{0.3cm} L_{\boldsymbol{b}_{v_2},\boldsymbol{d}_{v_2}}  \hspace*{0.3cm} L_{\boldsymbol{b}_{v_3},\boldsymbol{d}_{v_3}}  \hspace*{0.3cm}  L_{\boldsymbol{b}_{v_4},\boldsymbol{d}_{v_4}}]$ and $\boldsymbol{R} = [R_{\boldsymbol{d}_{c_1}}  \hspace*{0.3cm} R_{\boldsymbol{d}_{c_2}}  \hspace*{0.3cm} R_{\boldsymbol{d}_{c_3}}  \hspace*{0.3cm} R_{\boldsymbol{d}_{c_4}}  \hspace*{0.3cm} R_{\boldsymbol{d}_{c_5}}]$.}
	\label{Fig.MET_LDPC}
	%\vspace{-3em}
\end{figure}

\begin{figure}[!t]
	\centerline{
		\subfigure[]{\includegraphics[width=3in ]{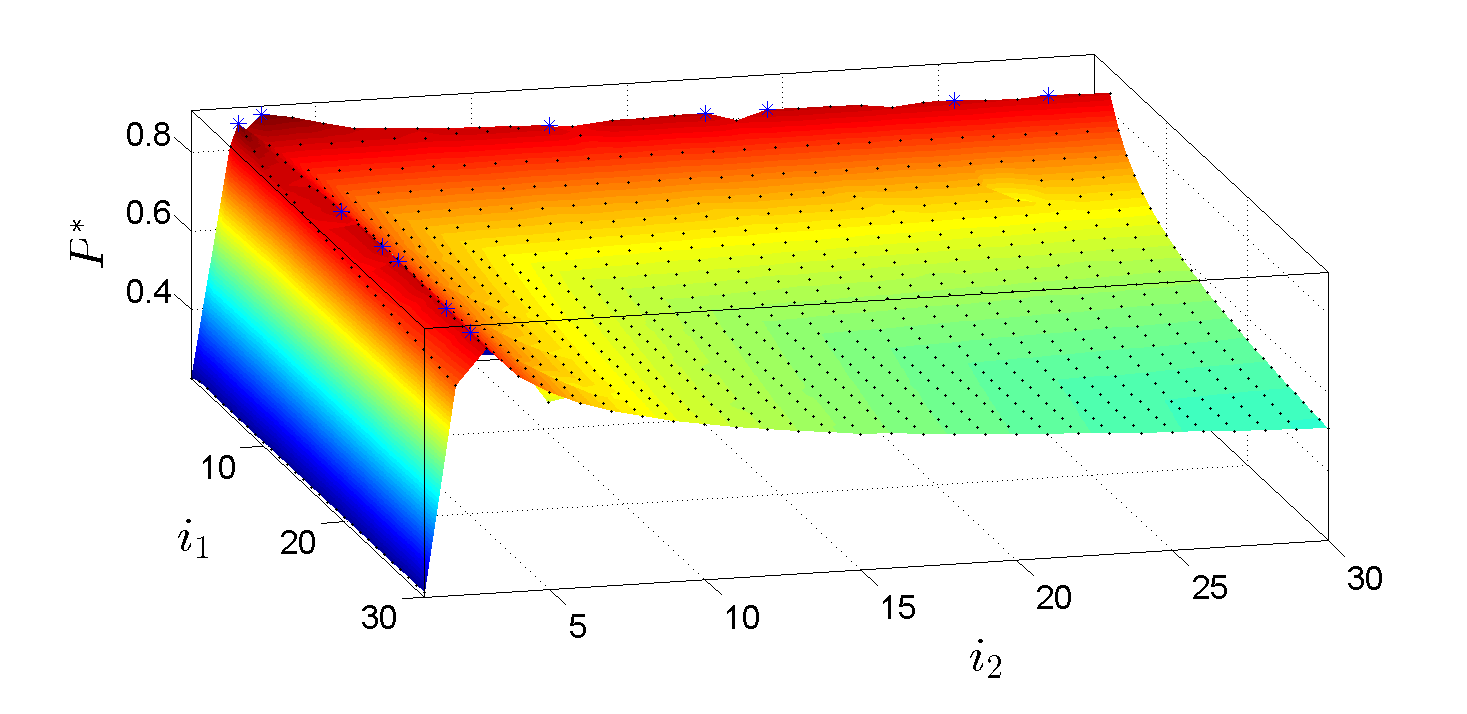}\label{rate_1_2_AWGN_outer} } 
		\hspace{1em}
		\subfigure[]{\includegraphics[width=3in]{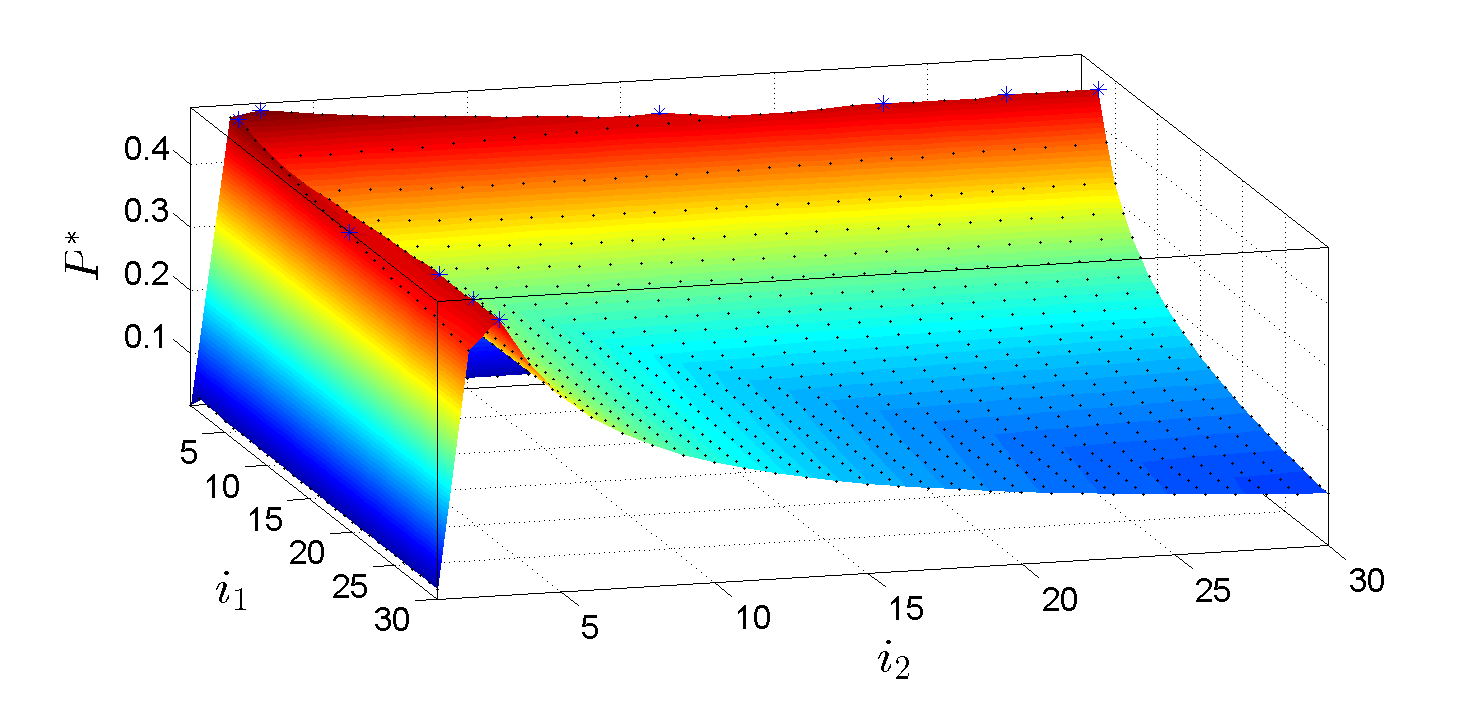}\label{rate_1_2_BEC_outer}} }
	\caption{The cost function for optimizing the rate-$1/2$ MET-LDPC code structure, where $\boldsymbol{\Lambda} = [([0~1], [i_1 ~0 ~0 ~0]) \hspace*{0.3cm} ([0~1], [i_2 ~0 ~0 ~0]) \hspace*{0.3cm} ([1~0], [0 ~3 ~3 ~0]) \hspace*{0.3cm} ([0~1], [0 ~0 ~0 ~1])]$. Blue colour \textquoteleft \textasteriskcentered \textquoteright  ~shows the locations of local maxima. }
	\fontsize{10}{1}
	\subcaption{(a)}{On the BI-AWGN channel}
	\subcaption{(b)}{On the BEC}
	\label{outer_cost_rate_1_2}
\end{figure}

\begin{figure}[!t]
	\centerline{
		\subfigure[]{\includegraphics[width=3in ]{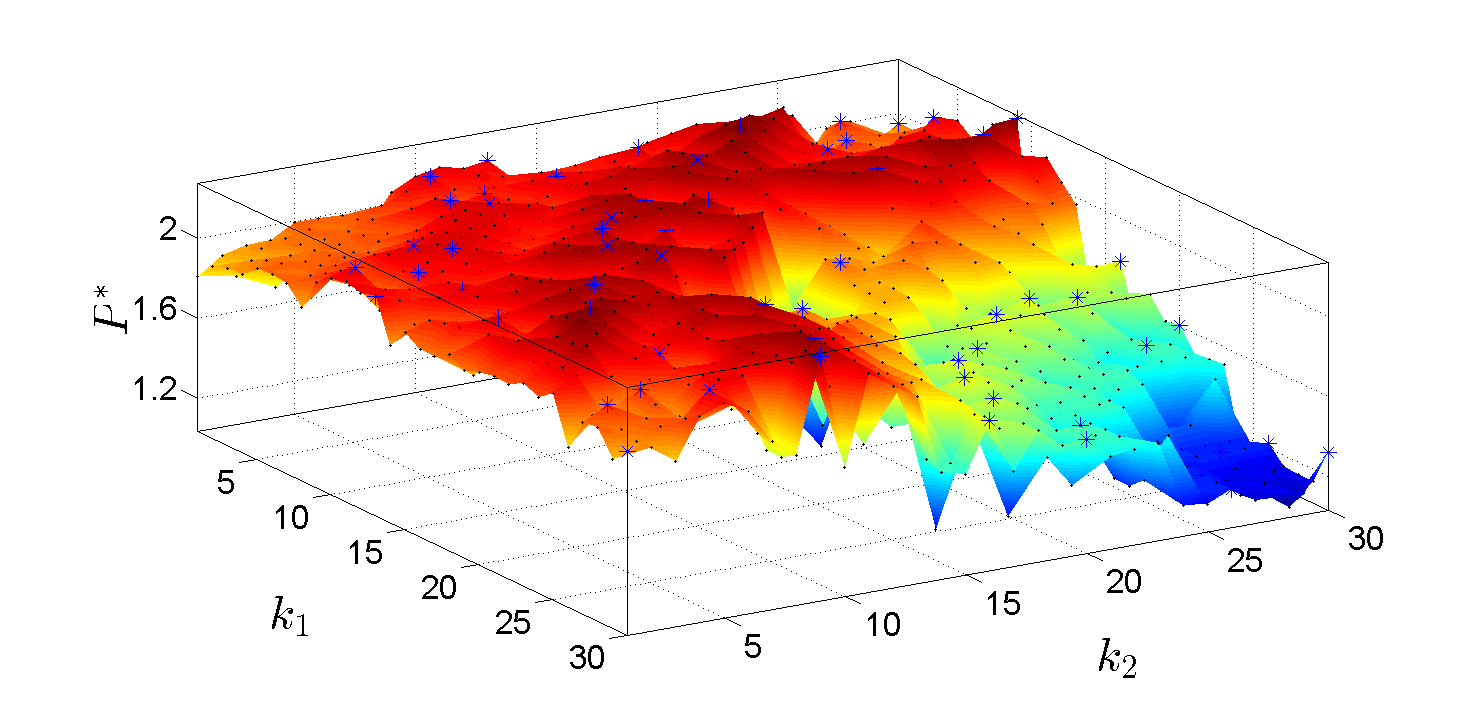} \label{rate_1_10_AWGN_outer}} 
		\hspace{1em}
		\subfigure[]{\includegraphics[width=3in]{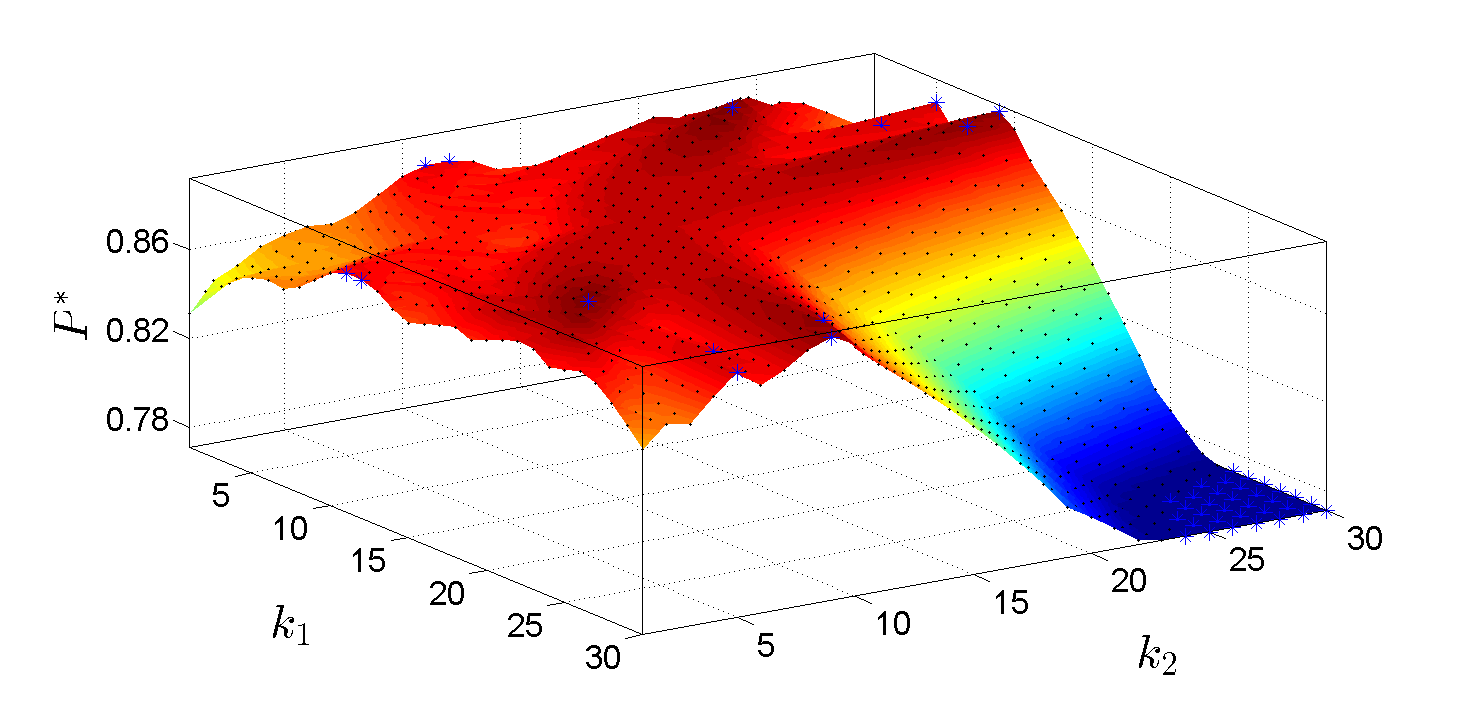}\label{rate_1_10_BEC_outer}} }
	\caption{The cost function for optimizing   the rate-$1/10$ MET-LDPC code structure, where $\boldsymbol{\Lambda} = [([0~1], [3 ~0 ~k_1 ~0]) \hspace*{0.3cm} ([0~1], [3 ~0 ~k_2 ~0])  \hspace*{0.3cm} ([0~1], [0 ~0 ~0 ~1])]$. Blue colour \textquoteleft \textasteriskcentered \textquoteright  ~shows the locations of local maxima. }
	\fontsize{10}{1}
	\subcaption{(a)}{On the BI-AWGN channel}
	\subcaption{(b)}{On the BEC}
	\label{outer_cost_rate_1_10}
\end{figure}

We then consider the cost function for optimizing  the Tanner graph structure, $\boldsymbol{\Lambda}$. In this case we use a four edge-type MET-LDPC code with four variable node classes (i.e., $v_{\max}  = 4$) and maximum allowable variable node degree of $30$  as shown in Fig.~\ref{Fig.MET_LDPC}. This corresponds to the same Tanner graph structure as the MET-LDPC code in Table VI of~\cite{RichardsonMulti2002}. For rate-$1/10$ MET-LDPC codes, we simply ignore the punctured variable node class in Fig.~\ref{Fig.MET_LDPC}  by making $v_{\max}  = 3$ which corresponds to the same Tanner graph structure as the MET-LDPC code in Table X of~\cite{RichardsonMulti2002}. The cost functions for this situation have multiple local maxima and they are far away from each other. Figs.~\ref{outer_cost_rate_1_2} and~\ref{outer_cost_rate_1_10} show  example  cost functions on the BI-AWGN channel and the BEC as the node degree of the class $1$ and class $2$ variable nodes are varied.

An another interesting feature noted is that, even though the maxima of  cost functions slightly differ from the BEC to the BI-AWGN channel,  both  have a similar shape. This suggests that we can use same code optimization techniques for the BEC and the BI-AWGN channel to optimize ($\boldsymbol{\Lambda}, \boldsymbol{\Gamma}, \boldsymbol{L}, \boldsymbol{R}$).

\subsection{Adaptive Range Method} \label{AR}
In this subsection, we briefly summarize the Adaptive Range (AR) method,  a type of local search  optimization technique, which   can alternatively be used to  optimize the degree distribution of MET-LDPC codes. The AR method forms the next set of points (arguments of the cost function) for evaluation as a random selection of the set of points close to the current optimum over the search space. The size of the search space is adapted as the algorithm progresses. 

The inputs of this algorithm are:
\begin{list}{-}{}
	\item Optimization parameters: Population size ($N_\text{P}$), range multiplier ($R_\text{M}$), search range ($S_\text{R}$), the tolerance on the best vector ($\delta$) 
	\item Tanner graph limits: $v_{\max},  c_{\max},  d_{v_{\max}},m_{e_{\max}}$
	%\item Decoder limits: Maximum number of decoder iterations ($\ell$)  
\end{list}
and the algorithm runs in following steps:
\begin{enumerate}	 
	\item {Initialization:
		For the first generation ($G=0$), choose $N_\text{P}$ length-$E$ vectors $\boldsymbol{Q}_{i,G}$, $i = 1, \dots ,N_\text{P}$, using the Queen’s move strategy~\cite{ShokrollahiDesignBook2005} where $E$ is the number of independent elements in ($\boldsymbol{\Lambda}, \boldsymbol{\Gamma}, \boldsymbol{L}, \boldsymbol{R}$).}
	\item{Threshold:
		For each vector $\boldsymbol{Q}_{i,G}$, run the DE algorithm  to calculate the ensemble threshold (${P}^{*}$). Then 
		select the vector with largest threshold ($\boldsymbol{Q}_{\text{Best},G}$, ${P}^{*}_{\text{Best},G}$) and the vector with next largest threshold ($\boldsymbol{Q}_{\text{NextBest},G}$, ${P}^{*}_{\text{NextBest},G}$).}
	\item {Random local search:
		For the next generation  $G+1$, new vectors are generated according to the following scheme. For each $\boldsymbol{Q}_{i,G+1}$, $i = 1, \dots ,N_\text{P}$ randomly choose each of its element, $\boldsymbol{Q}_{i,G+1}(j), j=1,\dots,E$  that is at most $S_\text{R}$ away from that of the current best vector. \vspace{-2em}							
		\begin{align*}
		\boldsymbol{Q}_{i,G+1}(j) &= \text{rand}[\max(\boldsymbol{Q}_{\text{Best},G}(j) - S_\text{R}),  
		\min(\boldsymbol{Q}_{\text{Best},G}(j) + S_\text{R})]\\
		\boldsymbol{Q}_{1,G+1}(j) &= \boldsymbol{Q}_{\text{Best},G}(j)
		\end{align*}}
	\item{Recalculation of search range:
		Recalculate  $S_\text{R}$ when ${P}^{*}_{\text{Best},G} - {P}^{*}_{\text{Best},G-1} < \delta$  \vspace{-1em}	
		\begin{align*}
		S_\text{R} &= R_\text{M} \times \max_{j \in 1..E} ( |\boldsymbol{Q}_{\text{Best},G}(j) - \boldsymbol{Q}_{\text{NextBest},G}(j) |)
		\end{align*}}
	\item{Stopping criterion:
		Halt if there is no improvement in threshold after three iterations.  Otherwise return to Step 2.}
\end{enumerate}

\section{Results and discussions}\label{discussion}
In this section, we present some numerical results obtained from the combined optimization algorithm. The MET-LDPC codes with rate-$1/2$ and rate-$1/10$ on the BEC and the BI-AWGN channel are considered. For the sake of comparison with existing optimization techniques for MET-LDPC codes, where only the degree distributions are optimized for a given Tanner graph structure, we also consider several Tanner graph structures from the literature and only optimized  the degree distribution. The decoding thresholds are computed using  DE approximation based on BER~\cite{LehmannAnalysisIT2003} and BEC density evolution algorithm~\cite{JohnsonIterativeBook2009} for the  BI-AWGN channel and the BEC, respectively.  All the algorithms (optimization and DE algorithms) are written in Matlab and run on an AMD Opteron(tm) 6386 SE  2.8
GHz PC. The results are presented in Tables~\ref{tab:rate1_2} to~\ref{tab:rate1_10_p} and  the table entries give the degree distribution  for the best case over 10 trials. 
Since we include $\boldsymbol{\Gamma}$ and $\boldsymbol{R}$ as  the dependent variables in the code optimization, we can calculate $\boldsymbol{\Gamma}=f(\boldsymbol{\Lambda}, \boldsymbol{L})$ and $\boldsymbol{R}=f'(\boldsymbol{\Lambda}, \boldsymbol{L})$, where $f(\cdot)$ and $f'(\cdot)$ are deterministic functions (see Section~\ref{check node}). $P^*_{\text{BEC}}$,  $P^*_{\text{AWGN}}$, $P_{\text{BEC}}$ and $P_{\text{AWGN}}$ denote the decoding threshold on the BEC,  the decoding threshold on the BI-AWGN channel, the Shannon limit for the BEC~\cite{richardson2001capacity} and the Shannon limit for the BI-AWGN channel~\cite{richardson2001capacity}, respectively. The numbers in brackets give the average for the decoding threshold and simulation time (CPU time) over the 10 trials. 

%As we explained Section~\ref{problem statemet}, there are two type of variables to be optimized i.e., the degree distribution ($\boldsymbol{L}, \boldsymbol{R}$) and the Tanner graph structure ($\boldsymbol{\Lambda}, \boldsymbol{\Gamma}$). Thus in this section we consider two different scenario: 1) Optimization of the MET-LDPC degree distribution ($\boldsymbol{L}, \boldsymbol{R}$), for a fixed Tanner graph structure. In this case  we select several Tanner graph structures from the literature and   ($\boldsymbol{L}, \boldsymbol{R}$) within a given Tanner graph structure is optimized. Thus there is no combined optimization involved. 2) We then consider our combined optimization methodology which optimizes both the Tanner graph structure and the degree distribution simultaneously (i.e., $\boldsymbol{\Lambda}, \boldsymbol{\Gamma}, \boldsymbol{L}, \boldsymbol{R}$) and demonstrate the benefits of the combined optimization method. 

\subsection{Optimization of the MET-LDPC degree distribution}\label{inner_optimzation}
We consider the optimization of the degree distribution of  MET-LDPC codes  for a fixed Tanner graph structure. 
The search for the best degree distribution for the MET-LDPC code ensemble is performed using the AR method described in Section~\ref{AR} as well as using the Dif.E~\cite{storn1997differential1997}. But in this work we present only the results obtained with AR method. This is because, similar to the optimization results shown in~\cite{JayasooriyaOptimizationITW2014}, we  found that  AR method and Dif.E returned the same optimal threshold up to three significant decimal places, and  the AR method is significantly quicker.  

We consider the MET-LDPC Tanner graph structures, $\boldsymbol{\Lambda}$, proposed in the literature, for rate-$1/2$~{\protect\cite[Table VI]{RichardsonMulti2002}} and rate-$1/10$~{\protect\cite[Table X]{RichardsonMulti2002}} MET-LDPC codes and only $\boldsymbol{L}$ within a given Tanner graph structure is optimized. We obtain $\boldsymbol{L}$ and $\boldsymbol{\Gamma}$ for a given ($\boldsymbol{\Lambda}, \boldsymbol{L}$) as per Section~\ref{check node}. In all cases, initialization was via the Queen’s move strategy~\cite{ShokrollahiDesignBook2005} and the stopping criterion was set as halt if there is no improvement in threshold after three generations and $R_\text{M}=1.25$ for the AR method. The population size at each generation was fixed to $100$.  The MET-LDPC degree distributions designed with the AR method for rate-$1/2$ and rate-$1/10$ are summarized in Tables~\ref{tab:rate1_2} and~\ref{tab:rate1_10}.

Comparing the asymptotic performance (i.e., decoding threshold) of the rate-$1/2$ MET-LDPC degree distribution optimized  for the BI-AWGN channel (code 2) with the reference rate-$1/2$ MET-LDPC (reference code 1,~{\protect\cite[Table VI]{RichardsonMulti2002}}), code 2 performs asymptotically within $0.054162$ from the Shannon limit, which is much better than the reference code 1. For the BEC, the decoding threshold value of the optimized  MET-LDPC degree distribution (code 1) is only $0.003394$ away from the Shannon limit. The MET-LDPC degree distributions designed for rate-$1/10$ on the BEC (code 5) and the BI-AWGN channel (code 6) perform  within $0.005225$ and $0.255808$ from the Shannon limit respectively, which are  closer than the MET-LDPC code proposed  in the literature~{\protect\cite[Table X]{RichardsonMulti2002}} (reference code 2).

% rate 1/2 result
\begin{table}[!t]
	\renewcommand{\arraystretch}{1.2}
	\processtable{Optimization of rate-$1/2$ MET-LDPC codes on the BEC and the BI-AWGN channel}
	{\tabcolsep6pt\begin{tabular}{@{}c| c |C{1.9cm} |C{1.9cm} C{1.9cm}| C{1.9cm} C{1.9cm} @{}}
		\cline{4-7}
		\multicolumn{2}{c}{}      & {} & \multicolumn{2}{c|}{Optimizing  ($\boldsymbol{L}$)} & \multicolumn{2}{c}{Optimizing  ($\boldsymbol{\Lambda},\boldsymbol{L}$)} \\
		\cline{4-7}
		\multicolumn{2}{c}{}      &   & \multicolumn{2}{c|}{AR} & \multicolumn{2}{c}{AR+Dif.E} \\
		\cline{3-7}
		\multicolumn{2}{c|}{}       & Reference code 1  & (Code 1) & (Code 2) & (Code 3) & (Code 4) \\
		\hline
		\multicolumn{2}{c|}{$\boldsymbol{\Lambda}$} &  \multicolumn{5}{c}{\multirow{2}[0]{*}{$\boldsymbol{L}$}} \\
		\cline{1-2}
		$\boldsymbol{b}$   & $\boldsymbol{d}$   & \multicolumn{5}{c}{} \\
		\hline
		\multicolumn{1}{c|}{[0 1]} & \multicolumn{1}{c|}{[2 0 0 0]} & 0.50  & 0.526258 & 0.300912 & 0.394302 & 0.346257 \\
		\multicolumn{1}{c|}{[0 1]} & \multicolumn{1}{c|}{[3 0 0 0]} & 0.30  & 0.124003 & 0.039927 & -     & - \\
		\multicolumn{1}{c|}{[0 1]} & \multicolumn{1}{c|}{[5 0 0 0]} & -     & -     & -     & -     & 0.037294 \\
		\multicolumn{1}{c|}{[0 1]} & \multicolumn{1}{c|}{[5 3 0 0]} & -     & -     & -     & 0.017512 & - \\
		\multicolumn{1}{c|}{[0 1]} & \multicolumn{1}{c|}{[0 0 0 1]} & 0.20  & 0.349739 & 0.659161 & 0.588186 & 0.616449 \\
		\multicolumn{1}{c|}{[1 0]} & \multicolumn{1}{c|}{[0 3 3 0]} & 0.20  & 0.271307 & 0.438600 & 0.389682 & 0.410453 \\
		\hline
		\multicolumn{2}{c|}{$\boldsymbol{\Gamma}=f(\boldsymbol{\Lambda},\boldsymbol{L})$} & \multicolumn{5}{c}{\multirow{2}[0]{*}{$\boldsymbol{R}=f'(\boldsymbol{\Lambda},\boldsymbol{L})$}} \\
		\cline{1-2}
		& $\boldsymbol{d}$   & \multicolumn{5}{c}{} \\
		\hline
		\multicolumn{1}{c|}{\multirow{11}[22]{*}{}} & \multicolumn{1}{c}{[2 4 0 0]} & -     & -     & 0.081393 & 0.028326 & 0.003028 \\
		\multicolumn{1}{c|}{} & \multicolumn{1}{c|}{[2 5 0 0]} & -     & -     & 0.035317 & -     & - \\
		\multicolumn{1}{c|}{} & \multicolumn{1}{c|}{[3 1 0 0]} & -     & 0.029215 & -     & -     & - \\
		\multicolumn{1}{c|}{} & \multicolumn{1}{c|}{[3 2 0 0]} & 0.10  & 0.232534 & -     & -     & - \\
		\multicolumn{1}{c|}{} & \multicolumn{1}{c|}{[3 4 0 0]} & -     & -     & -     & 0.257572 & 0.235631 \\
		\multicolumn{1}{c|}{} & \multicolumn{1}{c|}{[3 5 0 0]} & -     & -     & 0.162728 & 0.015598 & 0.055345 \\
		\multicolumn{1}{c|}{} & \multicolumn{1}{c|}{[4 1 0 0]} & 0.40  & -     & -     & -     & - \\
		\multicolumn{1}{c|}{} & \multicolumn{1}{c|}{[4 2 0 0]} & -     & 0.159819 & -     & -     & - \\
		\multicolumn{1}{c|}{} & \multicolumn{1}{c|}{[0 0 1 1]} & -     & -     & 0.002524 & 0.007326 & 0.001540 \\
		\multicolumn{1}{c|}{} & \multicolumn{1}{c|}{[0 0 2 1]} & -     & 0.235294 & 0.656637 & 0.580861 & 0.614910 \\
		\multicolumn{1}{c|}{} & \multicolumn{1}{c|}{[0 0 3 1]} & 0.20  & 0.114445 & -     & -     & - \\
		\hline
		\multicolumn{1}{l|}{\multirow{2}[0]{*}{Threshold }} & \multicolumn{1}{l|}{$P^*_{\text{BEC}}$} & \textbf{0.463135} & \textbf{0.496606} & -     & \textbf{0.497266} & - \\
		\multicolumn{1}{l|}{} & \multicolumn{1}{l|}{$P^*_{\text{AWGN}}$} & \textbf{0.895569} & -     & \textbf{0.924438} & -     & \textbf{0.927002} \\
		\hline
		\multicolumn{1}{l|}{Gap to the } & \multicolumn{1}{l|}{$|P_{\text{BEC}}-P^*_{\text{BEC}}|$} & 0.036865 & 0.003394 & -     & 0.002734 & - \\
		\multicolumn{1}{l|}{Shannon limit} & \multicolumn{1}{l|}{$|P_{\text{AWGN}}-P^*_{\text{AWGN}}|$} & 0.083031 & -     & 0.054162 & -     & 0.051598 \\
		\hline
		\multicolumn{2}{l|}{Average threshold} & -     & (0.495947) & (0.909009) & (0.496167) & (0.925037) \\
		\multicolumn{2}{l|}{Average CPU time (seconds)} & -     & (134.92) & (972.13) & (9484.75) & (13586.41) \\
		\hline
		\multicolumn{7}{c}{}
		\label{tab:rate1_2}%
	\end{tabular}}{}
\end{table}

% rate 1/10 result
\begin{table}[!t]
	\renewcommand{\arraystretch}{1.2}
	\processtable{Optimization of rate-$1/10$ MET-LDPC codes on the BEC and the BI-AWGN channel}
	{\tabcolsep6pt\begin{tabular}{@{}c| c |C{1.9cm} |C{1.9cm} C{1.9cm}| C{1.9cm} C{1.9cm} @{}}
		\cline{4-7}
		\multicolumn{2}{c}{}      & {} & \multicolumn{2}{c|}{Optimizing  ($\boldsymbol{L}$)} & \multicolumn{2}{c}{Optimizing  ($\boldsymbol{\Lambda},\boldsymbol{L}$)} \\
		\cline{4-7}
		\multicolumn{2}{c}{}      &   & \multicolumn{2}{c|}{AR} & \multicolumn{2}{c}{AR+Dif.E} \\
		\cline{3-7}
		\multicolumn{2}{c|}{}       & Reference code 2  & (Code 5) & (Code 6) & (Code 7) & (Code 8) \\
		\hline
		\multicolumn{2}{c|}{$\boldsymbol{\Lambda}$} &  \multicolumn{5}{c}{\multirow{2}[0]{*}{$\boldsymbol{L}$}} \\
		\cline{1-2}
		$\boldsymbol{b}$   & $\boldsymbol{d}$   & \multicolumn{5}{c}{} \\
		\hline
		\multicolumn{1}{c|}{[0 1]} & \multicolumn{1}{c|}{[0 2 21 0]} & -     & -     & -     & -     & 0.090209 \\
		\multicolumn{1}{c|}{[0 1]} & \multicolumn{1}{c|}{[1 2 20 0]} & -     & -     & -     & 0.021205 & - \\
		\multicolumn{1}{c|}{[0 1]} & \multicolumn{1}{c|}{[2 0 21 0]} & -     & -     & -     & 0.096380 & - \\
		\multicolumn{1}{c|}{[0 1]} & \multicolumn{1}{c|}{[3 0 16 0]} & -     & -     & -     & -     & 0.031507 \\
		\multicolumn{1}{c|}{[0 1]} & \multicolumn{1}{c|}{[3 0 20 0]} & 0.100 & 0.097046 & 0.095936 & -     & - \\
		\multicolumn{1}{c|}{[0 1]} & \multicolumn{1}{c|}{[3 0 25 0]} & 0.025 & 0.021940 & 0.022769 & -     & - \\
		\multicolumn{1}{c|}{[0 1]} & \multicolumn{1}{c|}{[0 0 0 1]} & 0.875 & 0.881013 & 0.881295 & 0.882415 & 0.878284 \\
		\hline
		\multicolumn{2}{c|}{$\boldsymbol{\Gamma}=f(\boldsymbol{\Lambda},\boldsymbol{L})$} & \multicolumn{5}{c}{\multirow{2}[0]{*}{$\boldsymbol{R}=f'(\boldsymbol{\Lambda},\boldsymbol{L})$}} \\
		\cline{1-2}
		& $\boldsymbol{d}$   & \multicolumn{5}{c}{} \\
		\hline
		\multicolumn{1}{c|}{\multirow{12}[24]{*}{}} & \multicolumn{1}{c|}{[4 8 0 0]} & -     & -     & -     & -     & 0.014058 \\
		\multicolumn{1}{c|}{} & \multicolumn{1}{c|}{[5 8 0 0]} & -     & -     & -     & -     & 0.000966 \\
		\multicolumn{1}{c|}{} & \multicolumn{1}{c|}{[5 9 0 0]} & -     & -     & -     & -     & 0.006692 \\
		\multicolumn{1}{c|}{} & \multicolumn{1}{c|}{[12 2 0 0]} & -     & -     & -     & 0.010345 & - \\
		\multicolumn{1}{c|}{} & \multicolumn{1}{c|}{[12 3 0 0]} & -     & -     & -     & 0.004298 & - \\
		\multicolumn{1}{c|}{} & \multicolumn{1}{c|}{[13 3 0 0]} & -     & -     & -     & 0.002943 & - \\
		\multicolumn{1}{c|}{} & \multicolumn{1}{c|}{[15 0 0 0]} & 0.025 & -     & -     & -     & - \\
		\multicolumn{1}{c|}{} & \multicolumn{1}{c|}{[18 0 0 0]} & -     & 0.003787 & -     & -     & - \\
		\multicolumn{1}{c|}{} & \multicolumn{1}{c|}{[19 0 0 0]} & -     & 0.015200 & 0.017989 & -     & - \\
		\multicolumn{1}{c|}{} & \multicolumn{1}{c|}{[20 0 0 0]} & -     & -     & 0.000717 & -     & - \\
		\multicolumn{1}{c|}{} & \multicolumn{1}{c|}{[0 0 2 1]} & -     & 0.153604 & 0.155935 & 0.199160 & 0.236358 \\
		\multicolumn{1}{c|}{} & \multicolumn{1}{c|}{[0 0 3 1]} & 0.875 & 0.727409 & 0.725360 & 0.683254 & 0.641927 \\
		\hline
		\multicolumn{1}{l|}{\multirow{2}[0]{*}{Threshold }} & \multicolumn{1}{l|}{$P^*_{\text{BEC}}$} &\textbf{ 0.876221} & \textbf{0.894775} & -     & \textbf{0.898315} & - \\
		\multicolumn{1}{l|}{} & \multicolumn{1}{l|}{$P^*_{\text{AWGN}}$} & \textbf{2.179504} & -     & \textbf{2.336792} & -     & \textbf{2.369385} \\
		\hline
		\multicolumn{1}{l|}{Gap to the } & \multicolumn{1}{l|}{$|P_{\text{BEC}}-P^*_{\text{BEC}}|$} & 0.023779 & 0.005225 & -     & 0.001685 & - \\
		\multicolumn{1}{l|}{Shannon limit} & \multicolumn{1}{l|}{$|P_{\text{AWGN}}-P^*_{\text{AWGN}}|$} & 0.413096 & -     & 0.255808 & -     & 0.223215 \\
		\hline
		\multicolumn{2}{l|}{Average threshold} & -     & (0.894775) & (2.205597) & (0.897388) & (2.353198) \\
		\multicolumn{2}{l|}{Average CPU time (seconds)} & -     & (174.98) & (40.18) & (4865.33) & (9686.83) \\
		\hline
		\multicolumn{7}{c}{}
		\label{tab:rate1_10}%
	\end{tabular}}{}
\end{table}

\subsection{Optimization of the MET-LDPC  Tanner graph structure and the degree distribution: The combined optimization}

Considering  both the degree distribution  and the Tanner graph structure  as the variables to be optimized, we searched for good MET-LDPC code ensembles using our proposed combined optimization method. We  represent the combined optimization method as (\textbf{A} + \textbf{B}) where \textbf{A} represents the optimization method use for optimizing $\boldsymbol{L}$, and \textbf{B}  for optimizing $\boldsymbol{\Lambda}$. We considered several different approaches for the combined optimization method: (AR+Dif.E), (AR+AR) and (Dif.E+Dif.E). Similar to the optimization results shown in~\cite{JayasooriyaOptimizationITW2014}, we found that (AR+Dif.E) optimize for the best MET-LDPC code ensemble (with optimal threshold) significantly quicker than other two methods. Thus in this paper we present only the results obtained with (AR+Dif.E). The initial setting for the  AR method is done as descried in Section~\ref{inner_optimzation}. The population size  at each generation was fixed at $50$ for  optimizing $\boldsymbol{L}$ and $10$ for  optimizing $\boldsymbol{\Lambda}$. 

To see how the combined optimization method  systematically find good  MET-LDPC  code ensembles with variety of rates and channels, we consider design of  rate-$1/2$ and rate-$1/10$ MET-LDPC code ensembles over the BEC and the BI-AWGN channel.  The results are presented in Tables~\ref{tab:rate1_2} and~\ref{tab:rate1_10} (last two columns).  
In the design of the presented rate-$1/2$ MET-LDPC codes, we have had some additional constraints, which allow for a fair comparison with the previously reported MET-LDPC codes in the literature~{\protect\cite[Table VI]{RichardsonMulti2002}}. We have limited the number of edge-types to four and avoid variable nodes with degree higher than $10$ (i.e., ${d}_{v_{\max}} = 10$). We have also  constrained the number of variable node classes  and check nodes classes to be $v_{\max} =4, c_{\max}=5$  respectively. Fig.~\ref{Fig.MET_LDPC} shows an example Tanner graph representation with these constraints. To  design rate-$1/10$ MET-LDPC codes, we simply ignore the punctured variable node class in Fig.~\ref{Fig.MET_LDPC}  by making $v_{\max}  = 3$,  and set ${d}_{v_{\max}} = 30$ which corresponds to the same Tanner graph structure as the MET-LDPC code in Table X of~\cite{RichardsonMulti2002}, thus allowing for a fair comparison. We then compare the designed MET-LDPC codes with the MET-LDPC codes reported in the literature based on the several criteria:  lower maximum variable node degree,  higher threshold and minimum gap to Shannon capacity limit.

Comparing the decoding threshold of the rate-$1/2$ MET-LDPC code optimized  for the BEC channel (code 3) with reference code 1~{\protect\cite[Table VI]{RichardsonMulti2002}},  code 3 performs asymptotically within $0.002734$ from the Shannon limit. %Code 3 also outperforms code 1, which we obtained by optimizing the degree distribution of Tanner graph structure of reference code 1.
For the BI-AWGN channel, the decoding threshold  of the  MET-LDPC code designed from the combined optimization (code 4)  is only $0.051598$ away from the Shannon limit. It is important to note that the code 4 gains this threshold improvement by having the same maximum variable node degree of six as the reference code 1.   The MET-LDPC codes designed for rate-$1/10$ on the BEC (code 7) and the BI-AWGN channel (code 8) perform  within $0.001685$ and $0.223215$ from the Shannon limit respectively, which are significantly closer than the reference code 2~{\protect\cite[Table X]{RichardsonMulti2002}} and the codes obtained from optimizing only the degree distribution (code 5 and code 6). Here, an interesting observation is the maximum variable node degree. More specifically,  we gain this threshold improvement by having maximum node degree of $23$ on the BI-AWGN channel, whereas the reference code 2  requires a maximum node degree of $28$ to obtain a threshold of $2.179504$.

\subsection{Remark on MET-LDPC code design}
In this section, we summarize some useful properties of MET code structure we observed in code optimization. 

The advantage of MET-LDPC codes is greater flexibility in Tanner graph structure which can significantly improve the decoding performance. We  see the benefits of the MET generalization  over standard LDPC code in our simulation results. Code 4 in Table~\ref{tab:rate1_2} achieves a threshold of ${P}^* = 0.927002$ on the BI-AWGN channel with a maximum variable node degree of six,  whereas  a good LPDC code require a maximum variable node degree of $30$ to obtain a threshold this high~\cite{ChungConstructionThesis2000}. The advantage of a code ensemble  with  lower degrees is that it helps to reduce the decoding complexity. At the same time its performance for small block length  codes is  good~\cite{RichardsonModernBook2008}.

%The results are quite encouraging. Compared to regular LDPC codes for which the highest achievable threshold for the BIAWGNC is , irregular LDPC codes have substantially higher thresholds. 

% rate 1/10 with punctured nodes
\begin{table}[!t]
	\renewcommand{\arraystretch}{1.2}
	%\processtable{The degree distributions of rate-$1/10$ MET-LDPC codes with punctured variable node designed using combined optimization (AR+Dif.E)}
	\caption{The degree distributions of rate-$1/10$ MET-LDPC codes with punctured variable node classes designed using combined optimization (AR+Dif.E)}
	\processtable{}
	{\tabcolsep6pt\begin{tabular}{@{}c| c |C{3cm} C{3cm}@{}}
		\cline{3-4}
		\multicolumn{2}{c|}{}        & (Code 9) & (Code 10)  \\
		\hline
		\multicolumn{2}{c|}{$\boldsymbol{\Lambda}$} & \multicolumn{2}{c}{\multirow{2}[0]{*}{$\boldsymbol{L}$}}  \\
		\cline{1-2}
		$\boldsymbol{b}$   & $\boldsymbol{d}$   & \multicolumn{2}{c}{} \\
		\hline
		\multicolumn{1}{c|}{[0 1]} & \multicolumn{1}{c|}{[2 0 5 0]} & -     & 0.274947 \\
		\multicolumn{1}{c|}{[0 1]} & \multicolumn{1}{c|}{[3 0 0 0]} & 0.019947 & 0.019292 \\
		\multicolumn{1}{c|}{[0 1]} & \multicolumn{1}{c|}{[4 2 0 0]} & 0.006135 & - \\
		\multicolumn{1}{c|}{[0 1]} & \multicolumn{1}{c|}{[0 0 0 1]} & 0.973919 & 0.705761 \\
		\multicolumn{1}{c|}{[1 0]} & \multicolumn{1}{c|}{[0 2 3 0]} & 0.721841 & - \\
		\multicolumn{1}{c|}{[1 0]} & \multicolumn{1}{c|}{[1 2 3 0]} & -     & 0.012436 \\
		\hline
		\multicolumn{2}{c|}{$\boldsymbol{\Gamma}=f(\boldsymbol{\Lambda},\boldsymbol{L})$} & \multicolumn{2}{c}{\multirow{2}[0]{*}{$\boldsymbol{R}=f'(\boldsymbol{\Lambda},\boldsymbol{L})$}} \\
		\cline{1-2}
		& $\boldsymbol{d}$   & \multicolumn{2}{c}{} \\
		\hline
		\multirow{9}[18]{*}{} & \multicolumn{1}{c|}{[0 2 0 0]} & 0.487816 & - \\
		& \multicolumn{1}{c|}{[0 3 0 0]} & 0.075729 & - \\
		& \multicolumn{1}{c|}{[1 3 0 0]} & 0.084378 & - \\
		& \multicolumn{1}{c|}{[3 0 0 0]} & -     & 0.181803 \\
		& \multicolumn{1}{c|}{[3 1 0 0]} & -     & 0.024692 \\
		& \multicolumn{1}{c|}{[4 1 0 0]} & -     & 0.000181 \\
		& \multicolumn{1}{c|}{[0 0 2 1]} & 0.756233 & 0.705240 \\
		& \multicolumn{1}{c|}{[0 0 3 1]} & 0.217686 & 0.000521 \\
		\hline
		\multicolumn{1}{l|}{\multirow{2}[0]{*}{Threshold }} & \multicolumn{1}{l|}{$P^*_{\text{BEC}}$} & \textbf{0.897949} & - \\
		& \multicolumn{1}{l|}{$P^*_{\text{AWGN}}$} & -     & \textbf{2.323975} \\
		\hline
		\multicolumn{1}{l|}{Gap to the}  & \multicolumn{1}{l|}{$|P_{\text{BEC}}-P^*_{\text{BEC}}|$} & 0.002051 & - \\
		\multicolumn{1}{l|}{Shannon limit} & \multicolumn{1}{l|}{$|P_{\text{AWGN}}-P^*_{\text{AWGN}}|$} & -     & 0.268625 \\
		\hline
		\multicolumn{4}{c}{}
		\label{tab:rate1_10_p}%
	\end{tabular}}{}
\end{table}

Another advantage of MET-LDPC code flexibility is that the MET framework allows for  punctured variable nodes in  code design. This is useful for lowering the maximum node degree of the code ensemble. We consider the design of rate-$1/10$ MET-LDPC codes  as it is difficult to design a good low-rate codes with a lower maximum node degree. The degree distribution of rate-$1/10$ MET-LDPC codes designed using combined optimization for the BEC and the BI-AWGN channel is shown in Table~\ref{tab:rate1_10_p}. Here we can see the benefit of adding punctured variable nodes to low-rate MET-LDPC codes.  For example, code 8 in Table~\ref{tab:rate1_10}, without punctured variable nodes, requires maximum variable node degree of $23$  to obtain a threshold of ${P}^* =2.369385$ on the BI-AWGN channel. However, we  obtain a very close threshold (${P}^* = 2.323975$) with the code 10 in Table~\ref{tab:rate1_10_p}, with maximum variable node degree of seven  by adding punctured variable nodes to the  MET code structure.

%\end{comment}
\section{Conclusion}\label{Conclusion}

This paper  proposed a joint optimization technique for MET-LDPC codes which allows the optimization of the MET Tanner graph structure and degree distribution given  the number of
edge-types and maximum node degrees.  This optimization method has allowed us to systematically find good  MET-LDPC  code structures as opposed to trail and error or intuition in conventional approaches. We found that this joint optimization, which uses the adaptive range method to optimize the degree distribution  and   differential evolution  to optimize the MET Tanner graph structure, is the most suitable approach for optimizing MET-LDPC codes.  In several examples, we demonstrated that MET-LDPC codes designed with the joint optimization method outperforms previously reported MET-LDPC codes.

\section{Acknowledgments}
This work is supported by the Australian Research Council under grants FT110100195, FT140100219, and DP150100903.

%\section{Appendices} \label{Appendices}

\end{document}